 \documentclass[twocolumn,trackchanges]{aastex631}

\usepackage{upgreek}

\shorttitle{\textit{TESS} albedos of hot Jupiters}
\shortauthors{Blažek et al.}
\graphicspath{{./}{figures/}}

\begin{document}

\title{Constraints on \textit{TESS} albedos for {f}ive hot Jupiters}

\correspondingauthor{Martin Blažek}
\email{deathlosopher@seznam.cz}

\author[0000-0002-6586-1907]{Martin Blažek}
\affiliation{Astronomical Institute of the Czech Academy of Sciences, Fri\v{c}ova 298, 251~65~~Ond\v{r}ejov, Czech Republic}
\affiliation{Department of Theoretical Physics and Astrophysics, Faculty of Science, Masaryk University, Kotl\'{a}\v{r}sk\'{a} 267/2, 611 37~~Brno, Czech Republic}

\author{Petr Kab\'{a}th}
\affiliation{Astronomical Institute of the Czech Academy of Sciences, Fri\v{c}ova 298, 251~65~~Ond\v{r}ejov, Czech Republic}

\author{Anjali A. A. Piette}
\affiliation{Institute of Astronomy, University of Cambridge, Madingley Road, Cambridge CB3 0HA, United Kingdom}
\affiliation{Earth and Planets Laboratory, Carnegie Institution for Science, 5241 Broad Branch Road, NW, Washington, DC 20015, USA}

\author{Nikku Madhusudhan}
\affiliation{Institute of Astronomy, University of Cambridge, Madingley Road, Cambridge CB3 0HA, United Kingdom}

\author{Marek Skarka}
\affiliation{Astronomical Institute of the Czech Academy of Sciences, Fri\v{c}ova 298, 251~65~~Ond\v{r}ejov, Czech Republic}
\affiliation{Department of Theoretical Physics and Astrophysics, Faculty of Science, Masaryk University, Kotl\'{a}\v{r}sk\'{a} 267/2, 611 37~~Brno, Czech Republic}

\author{J\'{a}n \v{S}ubjak}
\affiliation{Astronomical Institute of the Czech Academy of Sciences, Fri\v{c}ova 298, 251~65~~Ond\v{r}ejov, Czech Republic}
\affiliation{Astronomical Institute of Charles University, V Hole\v{s}ovi\v{c}k\'{a}ch 2, 180 00, Praha, Czech Republic}

\author{David R. Anderson}
\affiliation{Astrophysics Group, Keele University, Sta{f}fordshire ST5 5BG, United Kingdom}

\author{Henri M. J. Bo{f}{f}in}
\affiliation{European Southern Observatory, Karl-Schwarzschild-Str. 2, 85748 Garching bei München, Germany}

\author{Claudio C. C\'{a}ceres}
\affiliation{Departamento de Ciencias F\'{i}sicas, Facultad de Ciencias Exactas, Universidad Andr\'{e}s Bello, Av. Fernandez Concha 700, Las Condes, Santiago, Chile}
\affiliation{N\'ucleo Milenio de Formaci\'on Planetaria - NPF, Av. Gran Breta\~na 1111, Valpara\'{\i}so, Chile}

\author{Neale P. Gibson}
\affiliation{School of Physics, Trinity College Dublin, The University of Dublin, Dublin 2, Ireland}

\author{Sergio Hoyer}
\affiliation{Aix Marseille Universit\'{e}, CNRS, CNES, Laboratoire d’Astrophysique de Marseille UMR 7326, 13388, Marseille, France}

\author{Valentin D. Ivanov}
\affiliation{European Southern Observatory, Karl-Schwarzschild-Str. 2, 85748 Garching bei München, Germany}
\affiliation{European Southern Observatory, Ave. Alonso de Córdova 3107, Vitacura, Santiago, Chile}

\author{Patricio M. Rojo}
\affiliation{Departamento de Astronom\'{i}a, Universidad de Chile, Camino El Observatorio 1515, Las Condes, Santiago, Chile}



\begin{abstract}

Photometric observations of occultations of transiting exoplanets can place important constraints on the thermal emission and albedos of their atmospheres. We analyse photometric measurements and derive geometric albedo ($A_\mathrm{g}$) constraints for {f}ive hot Jupiters observed with \textit{TESS} in the optical: WASP-18\,b, WASP-36\,b, WASP-43\,b, WASP-50\,b and WASP-51\,b. For WASP-43\,b, our results are complemented by a VLT/HAWK-I observation in the near-infrared at $2.09\,\upmu$m. We derive the {f}irst geometric albedo constraints for WASP-50\,b and WASP-51\,b: $A_\mathrm{g}<0.445$ and $A_\mathrm{g}<0.368$, respectively. We {f}ind that WASP-43\,b and WASP-18\,b are both consistent with low geometric albedos ($A_\mathrm{g}<0.16$) even though they lie at opposite ends of the hot Jupiter temperature range with equilibrium temperatures of $\sim1400$\,K and $\sim2500$\,K, respectively. We report self-consistent atmospheric models which explain broadband observations for both planets from \textit{TESS}, \textit{HST}, \textit{Spitzer} and VLT/HAWK-I. We {f}ind that the data of both hot Jupiters can be explained by thermal emission alone and ine{f}{f}icient day-night energy redistribution. The data do not require optical scattering from clouds/hazes, consistent with the low geometric albedos observed.

\end{abstract}



\section{Introduction} \label{sec:intro}

Thermal emission observations of exoplanet atmospheres provide essential insights into their chemical compositions, thermal structures, energy transport and clouds/hazes \citep[e.g.,][]{Burrows2008b,Cowan2011,Parmentier2016,Madhusudhan2019}. In particular, optical and near-infrared occultation photometry allows the albedo (or re{f}lectance) of an exoplanet to be measured \citep[e.g.,][]{Cowan2011,Angerhausen2015,Esteves2015,Mallonn2019}. The albedo, in turn, provides key insights into the physical properties of the atmoshpere, including the presence of clouds and hazes \citep[e.g.,][]{Burrows2008b}. To study exoplanetary atmospheres, two measures of albedo are typically used. While the Bond albedo measures the fraction of stellar light re{f}lected over all wavelengths, the geometric albedo $A_\mathrm{g}$ is wavelength dependent. Speci{f}ically, the latter is used to describe the re{f}lectance of an atmosphere at optical wavelengths.

A high albedo is indicative of signi{f}icant optical scattering in the atmosphere and can therefore indicate the presence of clouds and/or hazes. To date, a range of albedo measurements have been made for exoplanetary atmospheres, suggesting clear to cloudy atmospheres. For example, several hot Jupiters have been found to have low albedos and are therefore thought to have little or no cloud coverage in the photosphere, e.g., TrES-2\,b ($A_\mathrm{g}=0.025$, \citealt{kipping}), WASP-12\,b ($A_\mathrm{g}<0.064$, \citealt{bell}), WASP-18\,b ($A_\mathrm{g}<0.048$, \citealt{shporer}). Meanwhile, several exoplanets across the mass range have been found to have larger albedos, suggesting more signi{f}icant clouds and/or hazes, e.g., HD\,189733\,b ($A_\mathrm{g}=0.40\pm0.12$, \citealt{Evans2013}), Kepler-7\,b ($A_\mathrm{g}=0.35\pm0.02$, \citealt{demory2011,demory2013}), HAT-P-11\,b ($A_\mathrm{g}=0.39\pm0.07$, \citealt{huber}), Kepler-10\,b ($A_\mathrm{g}<0.61$, \citealt{batalha}). Furthermore, phase-curve o{f}fsets observed in some exoplanets by the \textit{Kepler} space telescope \citep{borucki,demory2013,Angerhausen2015,Esteves2015,Shporer2015} suggest that clouds may be more prevalent in cooler planets, with a transition at $\sim1900$\,K between cloudy and non-cloudy atmospheres \citep{Parmentier2016}. Albedo measurements of hot Jupiters across a range of temperatures are therefore needed to further elucidate the presence of clouds and hazes across this regime.

Constraints on exoplanetary albedos also provide important information about the thermal properties of their atmospheres. Optical scattering from clouds and hazes cools the dayside, a{f}fecting the brightness temperatures measured in occultations \citep[e.g.,][]{Morley2013}. This can in turn a{f}fect inferences of day-night energy redistribution, as the cooling due to clouds/hazes may be degenerate with the e{f}fects of energy redistribution \citep{Cowan2011}. Previous studies of hot Jupiters have revealed typically low albedos \citep{Cowan2011,Angerhausen2015,Esteves2015,Mallonn2019}, and the Transiting Exoplanet Survey Satellite (\textit{TESS}; \citealt{ricker2015}) will provide valuable new constraints as it continues to expand the population of hot Jupiters with albedo measurements.

Near-infrared (NIR) and optical observations probe di{f}ferent atmospheric properties and are therefore highly complementary. In particular, the NIR probes thermal emission from exoplanet atmospheres and can place constraints on their chemical compositions and thermal pro{f}iles. The High Acuity Wide-{f}ield K-band Imager (HAWK-I) on the Very Large Telescope (VLT) probes the $\sim0.9$--2.4-$\upmu$m range and is well-suited to probing such thermal emission \citep[e.g.,][]{anderson, gibson2010}. Meanwhile, \textit{TESS} operates in the 0.6--1-$\upmu$m range and is ideally suited to search for re{f}lected light from exoplanet atmospheres \citep[e.g.,][]{shporer, beatty}. To date, \textit{TESS} has made con{f}irmed detections of over a hundred exoplanets, with more than a thousand detections currently awaiting con{f}irmation. While its primary goal is to search for new exoplanets orbiting bright stars, many occultations of already known exoplanets have been detected with \textit{TESS} phase curves \citep[e.g.,][]{shporer, bourrier2020}. The growing population of exoplanets with \textit{TESS} data is allowing comprehensive studies of atmospheric albedos across a range of exoplanets \citep{wong2020b}.

Our primary goal in this work is to constrain occultation depths, using observations from \textit{TESS} in the optical and from HAWK-I in the near-infrared, of these hot Jupiters: WASP-18\,b, WASP-36\,b, WASP-43\,b, WASP-50\,b, and WASP-51\,b. This in turn leads to constraints on the albedos of these planets, providing clues about their thermal properties, energy redistribution and clouds. We further use \textit{TESS} and HAWK-I data, in addition to existing \textit{Spitzer} data, to investigate atmospheric models for two hot Jupiters at opposite ends of the temperature range: WASP-43\,b \citep{hellier}, with an equilibrium temperature of $\sim1400$\,K, and WASP-18\,b \citep{southworth}, with an equilibrium temperature of $\sim2500$\,K. In particular, WASP-18\,b is at the transition between the \emph{hot} and \emph{ultra-hot} subcategories of hot Jupiters. This is an important regime as there can be signi{f}icant changes in atmospheric properties, including the thermal dissociation of molecules \citep[e.g.,][]{Arcangeli2018,Gandhi2020,Lothringer2018,parmentier2018b} and the presence of thermal inversions \citep[e.g.,][]{baxter2020}.

This article is structured as follows: in Section~\ref{sec:data} we describe the data sets and instrumentation, in Section~\ref{sec:analysis} we present data analysis, our results are presented in Section~\ref{sec:results}, and in Section~\ref{sec:atmospheres} we discuss atmospheric properties of the studied exoplanets.

\section{Observations} \label{sec:data}
We investigate occultation observations of hot Jupiters with two di{f}ferent facilities: \textit{TESS} in the optical from space and VLT HAWK-I in the near-infrared on the ground. The observations include datasets of {f}ive occultations observed with \textit{TESS} and one occultation observed with HAWK-I. In what follows, we describe these observations.

\subsection{Target selection}
We selected WASP targets for our study which were discovered in a scope of the WASP survey \citep{pollacco}. They include the following exoplanetary systems: WASP-18 \citep{hellier2009}, WASP-36 \citep{smith}, WASP-43 \citep{hellier}, WASP-50 \citep{gillon2011}, and WASP-51 \citep{johnson}.

The chosen targets were originally selected from unpublished (all but one) HAWK-I data in the ESO Science Archive. Usually, these targets were observed because of the expected larger, and thus favourable, occultation depth. We investigate if by using modern techniques such as Gaussian process-based methods, we could extract meaningful science from these neglected data, and to draw -- if possible -- some conclusion for future occultation observations. Due to the insu{f}{f}icient quality of the HAWK-I data to detect an occultation or to put meaningful upper limits, we further describe in this work only one HAWK-I archival data set -- WASP-43. The data are based on an observation made with ESO Telescope at the La Silla Paranal Observatory under programme ID 086.C-0222 (PI Micha\"el Gillon). The data set was used as a test benchmark for which occultation was published by \citet{gillon2012} and we re-analysed it with a di{f}ferent method.

Next, we mined the \textit{TESS} archive for observations of our original HAWK-I objects, and found that they all have been monitored between 2018 and 2021 in various \textit{TESS} sectors, so we used all the available data for this work.

The orbital and physical properties of all studied exoplanetary systems are listed in Table~\ref{tab:systems_info}.

\subsection{Instruments used to acquire the data sets}
The Transiting Exoplanet Survey Satellite \citep{ricker2015} contains four wide-angle 10-cm telescopes with associated CCDs working in the wavelength bandpass between 600 and 1000\,nm centred on 786.5\,nm. As \textit{TESS} observes brighter stars, the brightness of our targets is between 8.8 and 12.2\,mag in the optical \textit{TESS} band. Since the start of its operation in 2018, \textit{TESS} has been photometrically observing almost the whole sky in sectors, each covering a {f}ield of view $24\degr\times96\degr$. 

The instrument used to get the ground-based data described in this article is the High Acuity Wide-{f}ield K-band Imager (HAWK-I) at Very Large Telescope of ESO \citep{pirard,casali,kissler-patig,siebenmorgen}. It hosts six narrow-band {f}ilters and the {f}ield of view of HAWK-I is $7.5\arcmin\times7.5\arcmin$. The detector is composed of four chips, each of them with $2048\times2048$\,px and works in the near-infrared band between 0.85--2.50\,$\upmu$m. The pixel scale of HAWK-I is 0.1064 arcsec\,px$^{-1}$. For more details see HAWK-I User Manual\footnote{\url{https://www.eso.org/sci/facilities/paranal/instruments/hawki/doc.html}}.

\begin{table*}
	\centering
	\caption{Stellar characteristics and physical properties of the planetary systems analysed in this article. Here RA is the right ascension, DEC is the declination, $T_\mathrm{e{f}f,\star}$ is the e{f}fective temperature of the star, $m_\mathrm{TESS}$ is the apparent magnitude in the \emph{TESS} bandpass, $R_\mathrm{p}$ and $M_\mathrm{p}$ are the stellar radius and the mass, respectively, $a$ is the semi-major axis of the orbit, and $P$ is the orbital period of the planet. References: $^a$:~\citet{southworth}, $^b$:~\citet{smith}, $^c$:~\citet{gillon2012}, $^d$:~\citet{tregloan-reed}, $^e$:~\citet{enoch}. Note: WASP-51 corresponds to HAT-P-30.}
	\label{tab:systems_info}
	\begin{tabular}{lccccccccc}
		\hline
			System & RA ($\alpha$) & DEC ($\delta$) & Type & $T_\mathrm{e{f}f,\star}$ [K] & $m_\mathrm{TESS}$ [mag] & $R_\mathrm{p}~[\mathrm{R}_\mathrm{J}]$ & $M_\mathrm{p}~[\mathrm{M}_\mathrm{J}]$ & $a~[\mathrm{au}]$ & $P$ [d] \\
		\hline
			WASP-18$^a$ & $01^\mathrm{h}37^\mathrm{m}25^\mathrm{s}$ & $-45\degr40\arcmin40\arcsec$ & F6V & 6431 & ~~8.83  & 1.165 & 10.43 & 0.021 & 0.94 \\
			WASP-36$^b$ & $08^\mathrm{h}46^\mathrm{m}20^\mathrm{s}$ & $-08\degr01\arcmin37\arcsec$ & G2  & 5900 & 12.15 & 1.281 & 2.303 & 0.026 & 1.54 \\
			WASP-43$^c$ & $10^\mathrm{h}19^\mathrm{m}38^\mathrm{s}$ & $-09\degr48\arcmin23\arcsec$ & K7V & 4520 & 11.02 & 1.036 & 2.034 & 0.015 & 0.81 \\
			WASP-50$^d$ & $02^\mathrm{h}54^\mathrm{m}45^\mathrm{s}$ & $-10\degr53\arcmin53\arcsec$ & G9V & 5400 & 11.01 & 1.138 & 1.437 & 0.029 & 1.96 \\
			WASP-51$^e$ & $08^\mathrm{h}15^\mathrm{m}48^\mathrm{s}$ & $+05\degr50\arcmin12\arcsec$ & G0  & 6250 & ~~9.91 & 1.420 & 0.760 & 0.042 & 2.81 \\
		\hline
	\end{tabular}
\end{table*}

\subsection{Observations and data reduction}
While \textit{TESS} data are primarily intended to detect new exoplanets, here they serve as a probe of potential re{f}lected light in the optical wavelength range. The HAWK-I data set analysed here is the result of an observing run proposed to study atmospheres of highly irradiated transiting exoplanets. For the data reduction and to perform aperture photometry of this data set we used the Image Reduction and Analysis Facility (\textsc{iraf}; {\color{blue}Tody, D.}~\citeyear{tody1}, \citeyear{tody2}). While data reduction includes removing of e{f}fects of the used instrument which are added to raw images by the detector, aperture photometry includes summing the light of a given star in an aperture and subtracting sky background.

\subsubsection{\textit{TESS} full-phase data sets} \label{sec:tess_data_sets}
The available \textit{TESS} data of our targets were obtained in 2-minute cadence. The data sets analysed in this article were taken between August 2018 and March 2021, each of the sets comprises between roughly 13,000 and 18,000 data points and covers between 9 and 27 orbital phases (depending on the orbital period). The targets were observed in \textit{TESS} sectors 2--4, 7--9, 29--31, and 34--35.

To analyse the data, we used the Pre-search Data Conditioned Simple Aperture Photometry Flux, abbreviated as PDCSAP\_FLUX \citep{smith2012,stumpe2012,jenkins2016}, which is the {f}lux corrected for instrumental variations.

\subsubsection{HAWK-I occultation data set} \label{sec:hawki_sets}
Our HAWK-I data set was downloaded from the ESO archive. This data were obtained through a narrow-band {f}ilter of HAWK-I -- NB2090 ($2.09\,\upmu$m with width of 20\,nm). This data set of WASP-43 has already been previously analysed and published \citep{gillon2012}, using a Markov chain Monte Carlo algorithm \citep{gillon2010} to model the light curve. We selected this system as a benchmark for comparison of di{f}ferent {f}itting methods and for our re-analysis we used the Gaussian Processes method described in \citet{gibson2012}.

The data set was obtained in 2010 and consists of 184 science frames with integration time of 1.7\,s. Three comparison stars were observed along with the target star. Standard photometric data reduction using {f}lat-{f}ield frames was performed. Then di{f}ferential aperture photometry was performed and the star with the most stable {f}lux (TYC 5490-153-1) was used as a comparison star for the di{f}ferential photometry. The obtained data points were then binned per 2 minute time intervals.

During the observation, changes of meteorological conditions were as follows: humidity in a range 10--18\,per cent, seeing in a range $0.47$--$1.39\arcsec$, and airmass decreasing from 2.10 to 1.05 as the star on the sky was rising during the whole observation. 

\begin{figure}
	\includegraphics[width=\columnwidth]{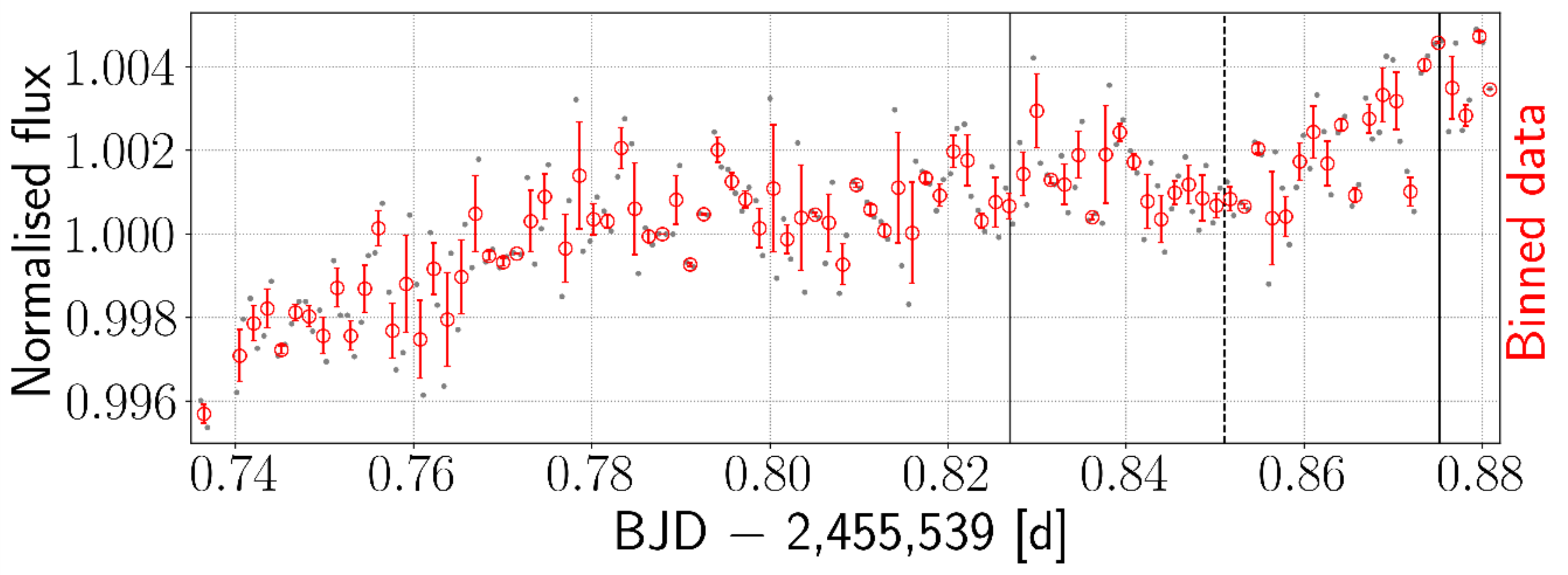}
	\caption{The raw light curve of WASP-43 system (HAWK-I, NB2090). The normalised raw light curve showing all data (the grey points) and data binned per 2 minutes (the red circles with error bars) is depicted. The vertical lines show the calculated beginning, the centre, and the end of the occultation.}
    \label{fig:w43_lcbin}
\end{figure}

The obtained light curve with the original data and the binned data is shown in Fig.~\ref{fig:w43_lcbin}. 

\section{Analysis of the photometric light curves} \label{sec:analysis}
In this section we describe the {f}itting methods used for all our data sets to derive occultation depths. We also present here the basic equations to theoretically estimate the occultation depth both from the re{f}lected light and from thermal emission.

\subsection{The {f}itting routines and detrending}
To {f}it the data sets, we used two di{f}ferent software packages. For {f}itting the \emph{TESS} phase curves, we used \texttt{allesfitter} package and for {f}itting the HAWK-I data set we used \texttt{GeePea} modelling routine. These two methods are described in the following subsections.

\subsubsection{`Allesfitter' software package}
To {f}it the \emph{TESS} phase curves, shown in Fig.~\ref{fig:allesPhaseCurves}, we used \textsc{python}-based \texttt{allesfitter} software package \citep{guenther2019,guenther2021}. It was developed to model photometric and radial velocity data. To make systematic noise models, Gaussian processes (GP) are included. After running the code an initial guess is obtained, then inference via MCMC or Nested Sampling is initiated. The methods include tests to assess convergence and also residual diagnostics to check possible structure in residuals. For details about \texttt{allesfitter} modelling package, see \citet{guenther2019,guenther2021} and the o{f}{f}icial website\footnote{\url{https://www.allesfitter.com/}}.

We {f}itted and sampled from the posterior of the ratio of the planetary to the stellar radius $R_\mathrm{p}/R_\star$, the sum of those radii divided by the semi-major axis $(R_\mathrm{p}+R_\star)/a$, cosine of the inclination angle of the planetary orbit $\cos{i}$, epoch, i.e., the time of the centre of the transit $T_0$, the ratio of the surface brightness of the planet to the star $J$, logarithm of the error scaling of white noise used for the GP $\ln\sigma$, a baseline o{f}fset $\Delta{F}$, the semi-amplitude of the Doppler-boosting $A_\mathrm{beaming}$, the amplitude of the atmospheric contribution (both thermal and re{f}lected) to the phase curve modulation $A_\mathrm{atmospheric}$, and the amplitude of the ellipsoidal modulation caused by tidal interaction between the host star and the planet $A_\mathrm{ellipsoidal}$. We {f}ixed the orbital period of the planet $P$, eccentricity and argument of periastron (planetary orbit) $\sqrt{e}\cos{\omega}$ and $\sqrt{e}\sin{\omega}$, and limb darkening coe{f}{f}icients $q_1$ and $q_2$. The values of $P, e$, and $\omega$ were adopted from discovery articles of the particular exoplanetary systems (Table~\ref{tab:systems_info}). To derive limb darkening coe{f}{f}icients we used the quadratic model of \texttt{PyLDTk} software package (\citealt{parviainen2015} describing the package and \citealt{husser2013} describing the spectrum library).

The derived parameters from our {f}its were the host star radius divided by the semi-major axis $R_\star/a$, the semi-major axis divided by the host star radius $a/R_\star$, the planetary radius divided by the semi-major axis $R_\mathrm{p}/a$, the planetary radius $R_\mathrm{p}$, the semi-major axis of the planetary orbit $a$, the inclination angle of the planetary orbit $i$, the transit and occultation impact parameter $b_\mathrm{tra}$ and $b_\mathrm{occ}$, the total and full-transit duration $T_\mathrm{tot}$ and $T_\mathrm{full}$, the epoch of the occultation $T_\mathrm{0;occ}$, the equilibrium temperature of the planet $T_\mathrm{eq;p}$, the transit and occultation depth $\delta_\mathrm{tra}$ and $\delta_\mathrm{occ}$, the nightside {f}lux of the planet $F_\mathrm{nightside;p}$, and host star density $\rho_\star$. Formulae of all derived parameters by \textsc{allesfitter} are listed in Table A3 of \citet{guenther2021}.

We used all the \emph{TESS} photometric data of the systems available to date. Particularly, for WASP-18 modelling we used data of sectors 2, 3, 29, and 30, for WASP-36 data of sectors 8 and 34, for WASP-43 data of sectors 9 and 35, for WASP-50 data of sectors 4 and 31, and for WASP-51 data of sectors 7 and 34. For each sector of every system we period-folded the light curves and then merged all the light curves of each system together. Finally, we binned the data sets per 5-minute time intervals.

For each modelling, we used both MCMC and Nested Sampling method to {f}it our data and to derive parameters and their uncertainties. Both the methods perfectly agreed and gave results with negligible di{f}ferences. As Nested Sampling ensures that all convergence criteria are ful{f}illed, we present only the results obtained from this method (Section~\ref{sec:upper_limits_tess}).

\subsubsection{`GeePea' routine}
\label{sec:geepea'}
For {f}itting our HAWK-I data set (WASP-43), we used a Gaussian Processes method. The method is de{f}ined as an in{f}inite set of Gaussian variables which have common Gaussian distribution. The systematics are modelled here as a stochastic process. The GP model, our eclipse model, is a set of a deterministic component and a stochastic component. These are represented here as a mean function (the light-curve model) and a kernel function (the noise model), respectively. To implement our GP, we used the \texttt{GeePea} code\footnote{Available at \url{https://github.com/nealegibson}} as described in {\color{blue}Gibson et al.}~(\citeyear{gibson2013a}, \citeyear{gibson2013b}) and \citet{gibson2014}.

If we model a light curve of transit or occultation by using the GP, we have to assign parameters to the mean and kernel function (we will refer the parameters of the kernel function to `hyperparameters'). The kernel function takes at least three hyperparameters: height scale $\xi$ physically representing the typical range of the data points on the y-axis, a vector of length scale parameters $\eta$ physically representing changes on the x-axis (distance between `bumps'), and white noise $\sigma^2_\mathrm{w}$. We assign an array of parameters to the mean function, which represents the light-curve model. These parameters are time of the occultation centre $T_0$, orbital period $P$, scaled semi-major axis $a/R_\star$, planet-star radii ratio $R_\mathrm{p}/R_\star$, impact parameter $b$, out-of-transit {f}lux $f_\mathrm{oot}$, time gradient $T_\mathrm{grad}$, expected occultation depth $\delta_\mathrm{occ}$ and, in the case of a primary transit, also limb darkening coe{f}{f}icients $q_1$ and $q_2$. As the HAWK-I data are obtained only during the planetary occultation, we {f}itted only the occultation. 

Before the run of the {f}itting routine, $P$, $a/R_\star$, $R_\mathrm{p}/R_\star$, and $b$ were taken from literature and thus {f}ixed. The {f}itted parameters were $T_{0}$, $f_\mathrm{oot}$, $T_\mathrm{grad}$, $\delta_\mathrm{occ}$, and hyperparameters of the kernel function $\xi$, $\eta$, and $\sigma^2_\mathrm{w}$. In our case, besides time ($\eta_t$), we used airmass ($\eta_a$) as the second component of the length scale vector $\eta$.

To detrend the {f}itted light curve, we used polynomial regression of degree two assuming the out-of-occultation model to be a quadratic function of time ($f(t) = at^2 + bt + c$). For the polynomial regression we excluded data during the occultation. After inferring their values we calculated the function $f(t)$ for all the data points. To get the detrended and normalised-to-one {f}lux and the occultation model, we subsequently divided our data by the the polynomial function. 

We describe results of the HAWK-I light curve {f}it of WASP-43 in Section~\ref{sec:upper_limits_hawki}.

\subsection{Occultation depth estimation} \label{sec:ODE}
One of the input parameters of the \texttt{GeePea} routine is an estimated {f}lux drop during the occultation searched in our data which is then re{f}ined by the routine. The value is also needed to interpret the data and compare it with atmospheric models. To get the {f}lux drop estimation, we used a formula to calculate the occultation depth caused by re{f}lected light \citep[by a Lambert surface, i.e., a surface which scatters intensity isotropically; e.g.,][]{winn_book}:
\begin{equation}
    \delta_\mathrm{occ,re}=A_\mathrm{g}
    \left(\frac{R_\mathrm{p}}{a}\right)^2, 
    \label{eq:deltaOccRe}
\end{equation}
where $A_\mathrm{g}$ is the wavelength dependent geometric albedo (ratio of the {f}lux of a planet at full phase to the {f}lux of a perfectly di{f}fusing Lambert disc), $R_\mathrm{p}$ is the planetary radius and $a$ is the semi-major axis of the orbit. For putting upper limits on occultation depths we assume $A_\mathrm{g}$ equal to one which sets the maximum possible value of the occultation depth due to re{f}lected light.

During an occultation, the radiation {f}lux of the system is decreased as the thermal radiation from the planet is no longer seen while the planet is behind the star. To include that, we used the following formulation to estimate the thermal contribution of the planet:
\begin{equation}
    \delta_\mathrm{occ,th}=\left(\frac{R_\mathrm{p}}{R_\star}\right)^2 \frac{B_\lambda(T_\mathrm{p})}{B_\lambda(T_\star)},
    \label{eq:deltaOccTh}
\end{equation}
where $R_\star$ is the stellar radius and $B_\lambda$ ($T_\mathrm{p},T_\star$) are the Planck's functions corresponding to temperatures of the planet ($T_\mathrm{eq}$) and the star ($T_\mathrm{e{f}f,\star}$), approximating them as blackbody radiators. 

\subsection{Estimation of temperatures}
To estimate the equilibrium temperature of a planet we used this formula:
\begin{equation}
    T_\mathrm{eq,p}=T_\mathrm{e{f}f,\star} \sqrt{\frac{R_\star}{a}}\sqrt[4]{f(1-A_\mathrm{B})}.
    \label{eq:teq}
\end{equation}
Here $T_\mathrm{e{f}f,\star}$ is the e{f}fective temperature of the parent star, $R_\star$ is its radius, $A_\mathrm{B}$ is the Bond albedo (including radiation at all frequencies scattered into all directions), and $f$ is a {f}lux correction factor connected with redistribution of the stellar radiation over the planet's hemispheres.

Knowing the occultation depth from our {f}it and approximating planets and stars to be blackbody radiators, the brightness temperature $T_\mathrm{b}$ can be calculated from the occultation depth $\delta_\mathrm{occ}$ as follows:
\begin{equation}
	T_\mathrm{b} (\lambda) = \left(\frac{\mathrm{hc}}
	{\lambda \mathrm{k}_\mathrm{B}}\right) \left\lgroup \ln \left(1 + \frac{2\mathrm{h} \mathrm{c}^2k^2} {\delta_\mathrm{occ}\lambda^5 B_{\lambda,\star} }\right)\right\rgroup^{-1},
		\label{eq:tb}
\end{equation}
where h is Planck's constant, c is the speed of light in vacuum, k$_\mathrm{B}$ is the Ste{f}fan-Boltzmann constant, $\lambda$ is the wavelength at which we observed, $\delta_\mathrm{occ}$ is the measured occultation depth, and $B_{\lambda,\star}$ is the Planck function corresponding to $T_\mathrm{e{f}f}$ of the star. We have also denoted $k^2 \equiv \left(R_\mathrm{p}/R_\star\right)^2$ and used the wavelength dependent form of the Planck's law.

\section{Results} \label{sec:results}
Following the methods described in Section~\ref{sec:analysis}, in Section~\ref{sec:upper_limits_tess} we describe our results for the \emph{TESS} phase curves for each hot Jupiter target and in Section~\ref{sec:upper_limits_hawki} for the HAWK-I occultation for WASP-43\,b. 

\subsection{\textit{TESS} phase curve models and upper limits} \label{sec:upper_limits_tess}
We were able to detect primary transits of all the systems in the \textit{TESS} data sets. We have also detected the occultation of WASP-18\,b, which has the brightest host star among the systems we consider here. For the other systems we were able to place upper limits on their occultation depths and corresponding upper limits on their geometric albedos. For each binned data set we have also calculated the standard deviation of the weighted mean (RMS$_\mathrm{w}$), serving as a measure of the quality of the data set and which can also be compared with expected and derived occultation depths.

To derive $3\sigma$ upper limits on the occultation depths of WASP-36\,b, WASP-43\,b, WASP-50\,b, and WASP-51\,b, we took values of the upper and the lower uncertainties of the derived occultation depth, averaged them, and multipied by three (i.e., $3[(\sigma_+ + \sigma_-)/2]$). A corresponding upper limit on the geometric albedo can be obtained using Equation~\ref{eq:deltaOccRe} and by estimating the contribution of re{f}lected light, $\delta_\mathrm{occ,re}$, to the observed occultation depth. To do this, we use Equation~\ref{eq:deltaOccTh} to estimate the contribution of thermal emission to the occultation depth, and subtract it from the observed occultation depth: $\delta_\mathrm{occ,re}=\delta_\mathrm{occ}-\delta_\mathrm{occ,th}$.   

\begin{figure}
	\includegraphics[width=\columnwidth]{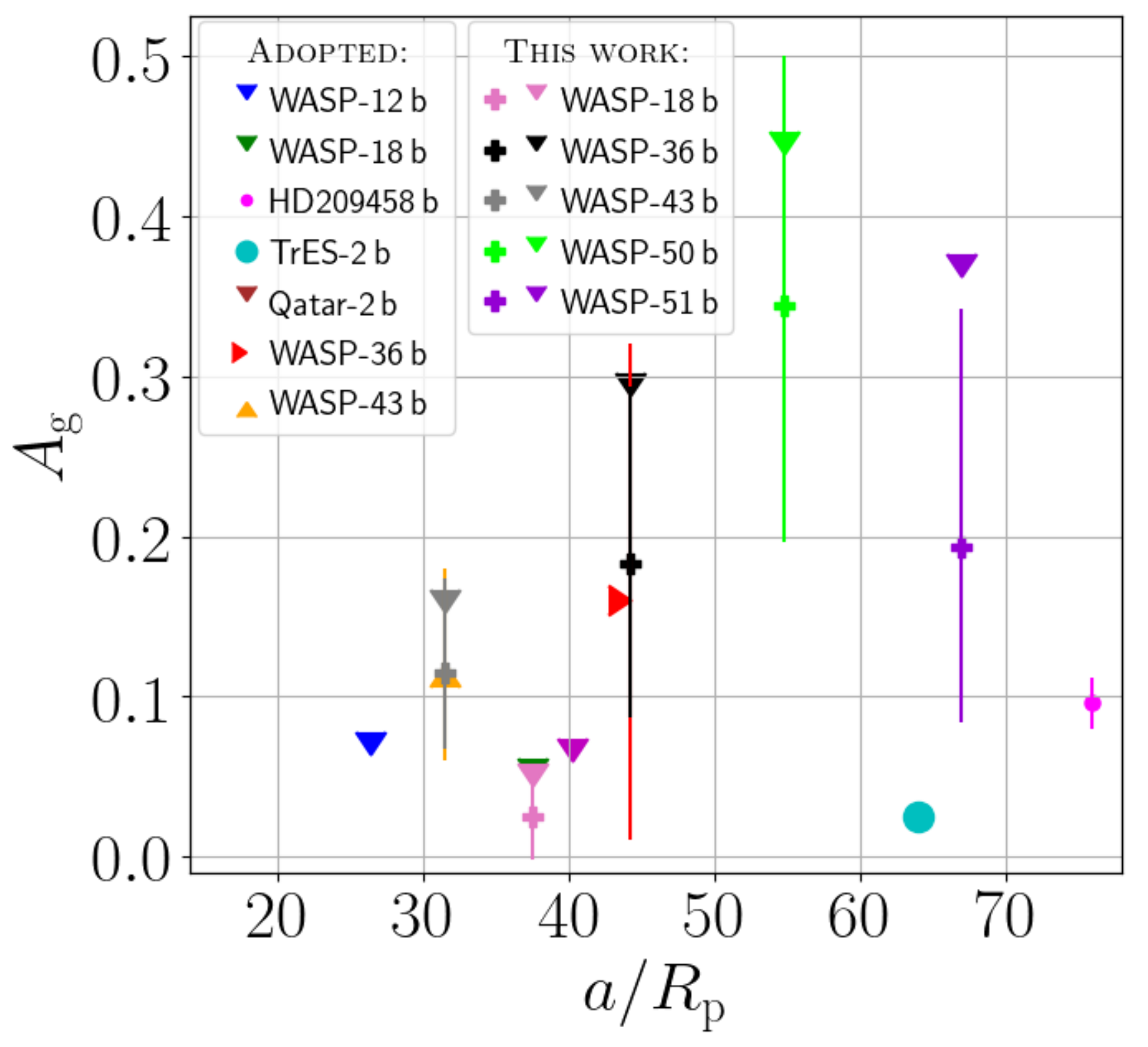}
		\caption{Optical geometric albedo as a function of the semi-major axis to planetary radius ratio, $a/R_\textrm{p}$. We include albedo constraints from the literature for seven hot Jupiters (`\textsc{Adopted}') as well as the \textit{TESS} albedos derived here for WASP-18\,b, WASP-36\,b, WASP-43\,b, WASP-50\,b, and WASP-51\,b (`\textsc{This work}'). The downward triangles ($\blacktriangledown$) show upper limits while a di{f}ferent symbol (bold `+' for our results) with an error bar shows a derived value including uncertainties. Literature references and corresponding instruments used are: WASP-12\,b: \citet{bell}, \emph{HST} STIS; WASP-18\,b: \citet{shporer}, \emph{TESS}; HD\,209458\,b: \citet{brandeker2022}, \emph{CHEOPS}; TrES-2\,b: \citet{kipping}, \emph{Kepler}; Qatar-2\,b: \citet{dai}, \emph{K2 (Kepler)}; WASP-36\,b \& WASP-43\,b: \citet{wong}, \emph{TESS}.}
		\label{fig:albVSarp}
\end{figure}

In Table~\ref{tab:upper_limits}, we summarise our constraints on the transit/occultation depths and geometric albedos of each planet. We also show `expected' values of the occultation depth for each planet ($\delta_\mathrm{occ,exp}$), calculated as the sum of the `expected' occultation depths due to re{f}lected light ($\delta_\mathrm{occ,re,exp}$) and thermal emission  ($\delta_\mathrm{occ,th,exp}$). These contributions are de{f}ined by Equations~\ref{eq:deltaOccRe} and \ref{eq:deltaOccTh}, respectively, assuming a limiting case of $A_\mathrm{g}=1$ and $T_\mathrm{p}=T_\mathrm{eq,p}$, where $T_\mathrm{eq,p}$ is de{f}ined according to Equation~\ref{eq:teq} with $f=0.25$ and $A_\mathrm{B}=0$. For all {f}ive hot Jupiters considered here, the occultation depth constraint (whether a detection or an upper limit) is lower than the `expected' occultation depth due to re{f}lection alone, $\delta_\mathrm{occ,re,exp}$. This indicates that $A_\mathrm{g}<1$ for these planets, as expected given existing constraints on hot Jupiter albedos \citep[e.g.,][]{Esteves2015}. Fig.~\ref{fig:albVSarp} shows our derived geometric albedo constraints as a function of $a/R_\mathrm{p}$ for all the planets studied in this work, alongside existing optical albedo constraints from the literature. The ratio $a/R_\mathrm{p}$ can be used to identify how well an exoplanet {f}its the characteristics of a hot Jupiter; a lower value means that the planet is closer to its parent star and/or has a larger radius.

\begin{figure}
   \includegraphics[width=\columnwidth]{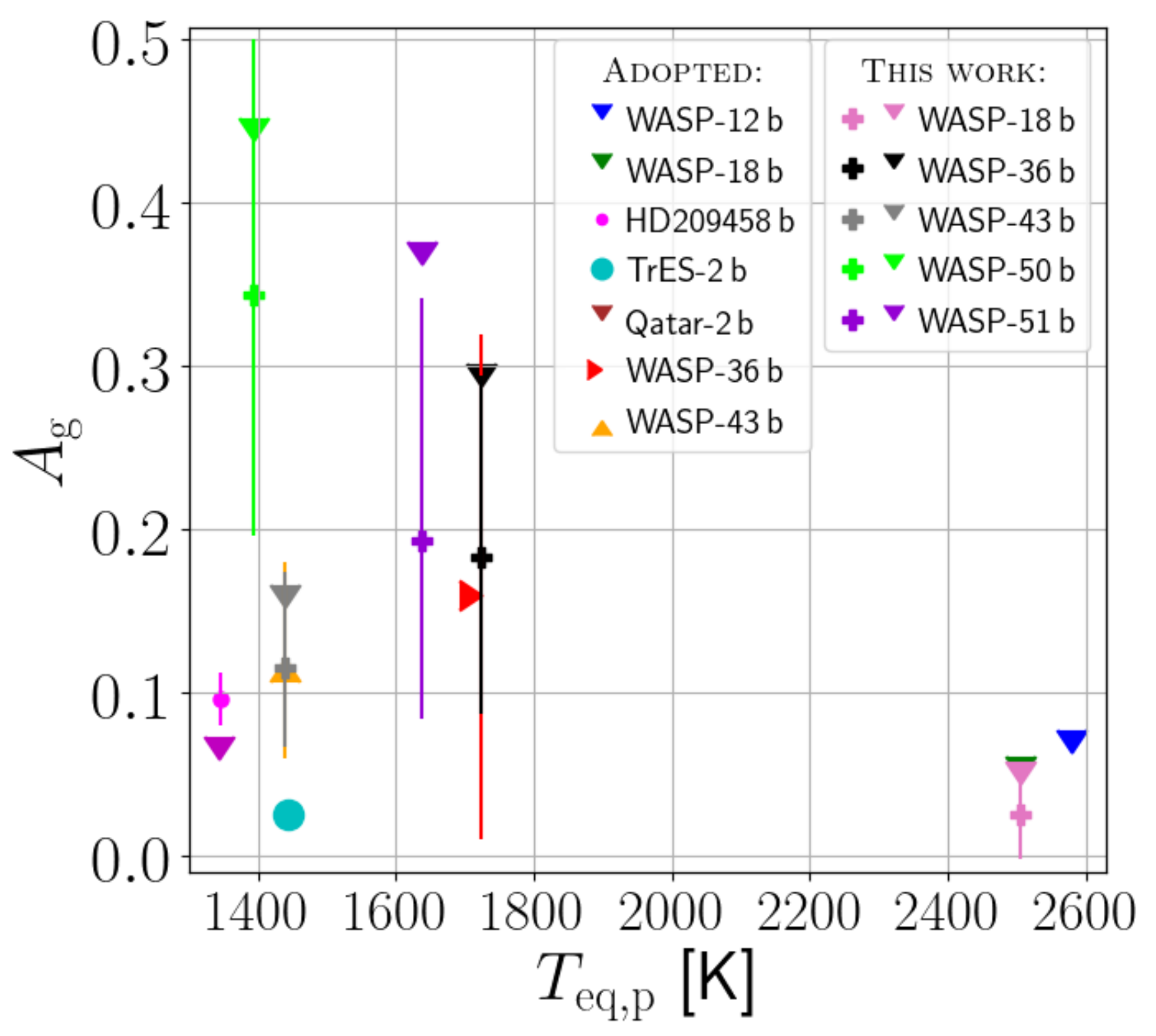}
        \caption{Optical geometric albedo as a function of equilibrium temperature, $T_\textrm{eq,p}$, for the same planets as in Fig.~\ref{fig:albVSarp}. Meaning of the used symbols, literature references and instruments used are the same as in Fig.~\ref{fig:albVSarp}.}
    \label{fig:albVStemp}
\end{figure}

In Fig.~\ref{fig:albVStemp}, we show geometric albedo as a function of equilibrium temperature for the same planets as in Fig.~\ref{fig:albVSarp}. The geometric albedo upper limits which we derive in this work for WASP-18\,b, WASP-36\,b, WASP-43\,b, WASP-50\,b, and WASP-51\,b all lie below 0.45. This is consistent with previous works which {f}ind that hot Jupiters typically have low albedos \citep[e.g.,][]{heng2013,Esteves2015,Mallonn2019,brandeker2022}, though higher optical albedos have also been measured in some cases \citep[e.g.,][]{Esteves2015,Niraula2018,wong,Adams2021,Heng2021}. The geometric albedos shown in Fig.~\ref{fig:albVStemp} are consistent with a range of values, with upper limits spanning $\lesssim0.05$ to $\sim0.45$. The diversity seen in hot Jupiter albedos may be indicative of a variety of cloud types and processes \citep{Adams2021}. Future albedo measurements spanning a wider range of equilibrium temperatures will be needed to further elucidate the nature of optical scattering in hot Jupiter atmospheres.

\begin{figure*}
	\includegraphics[height=7.8cm]{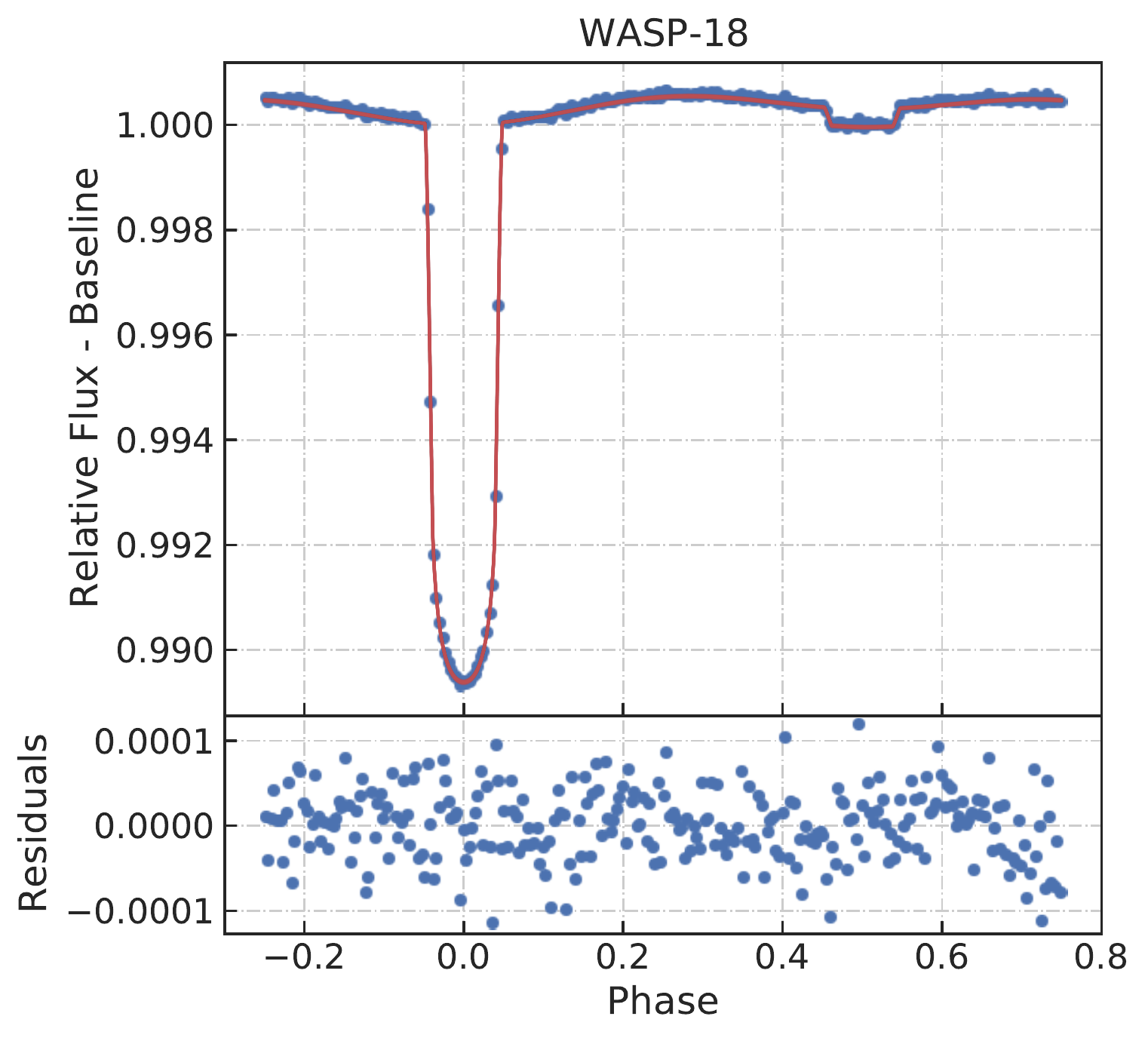}
	\includegraphics[height=7.8cm]{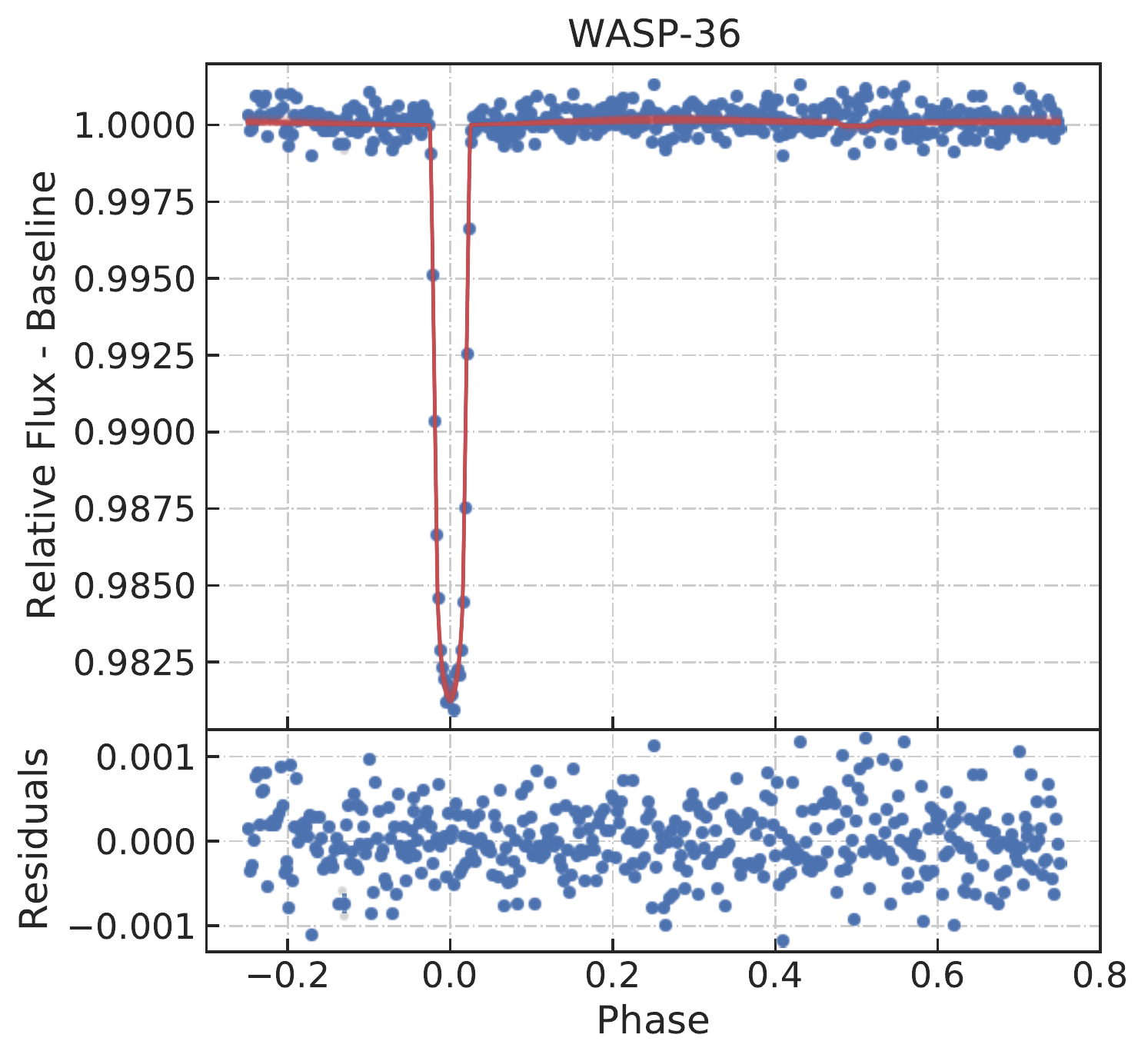}\\
	\includegraphics[height=7.8cm]{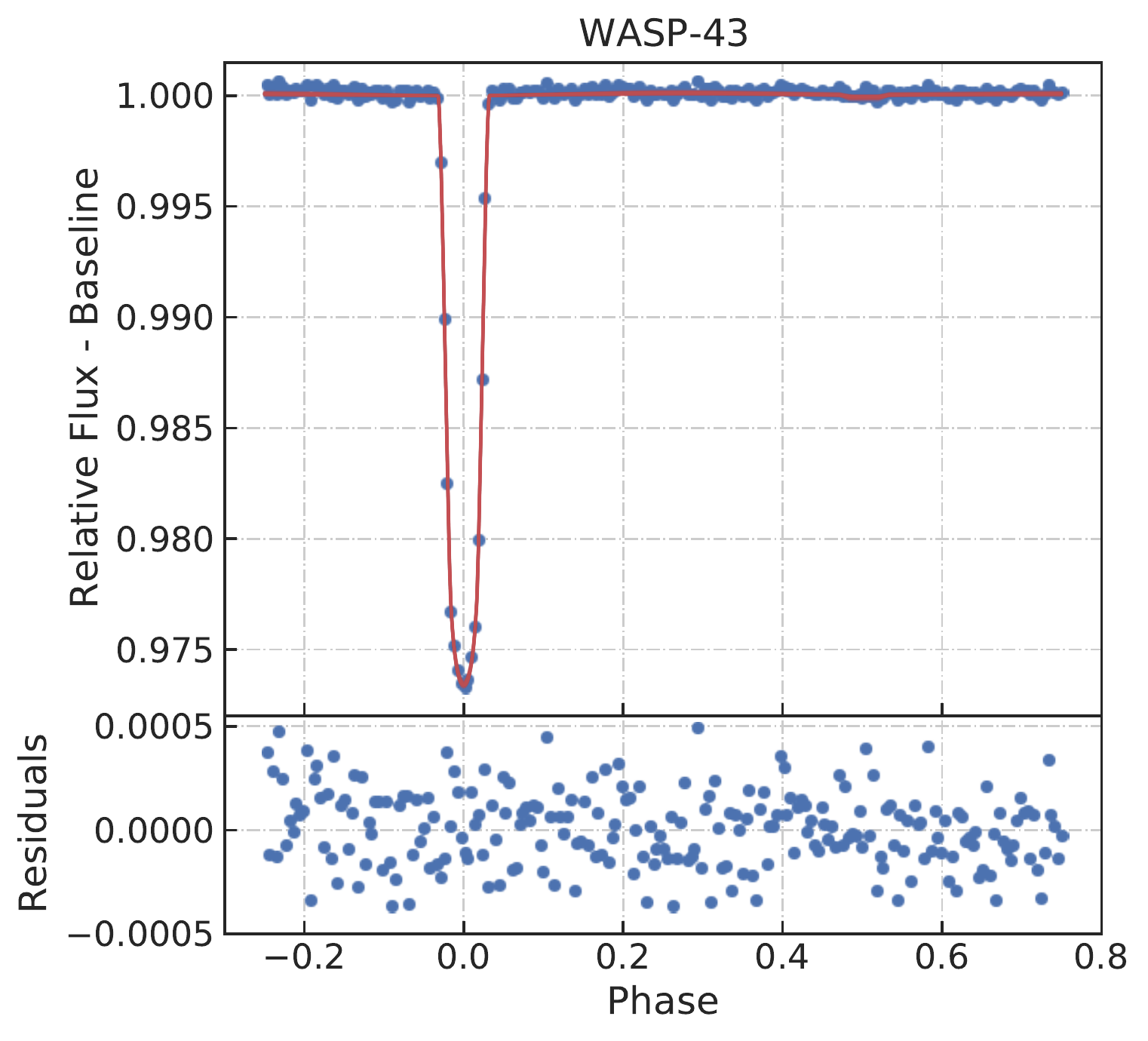}
	\includegraphics[height=7.8cm]{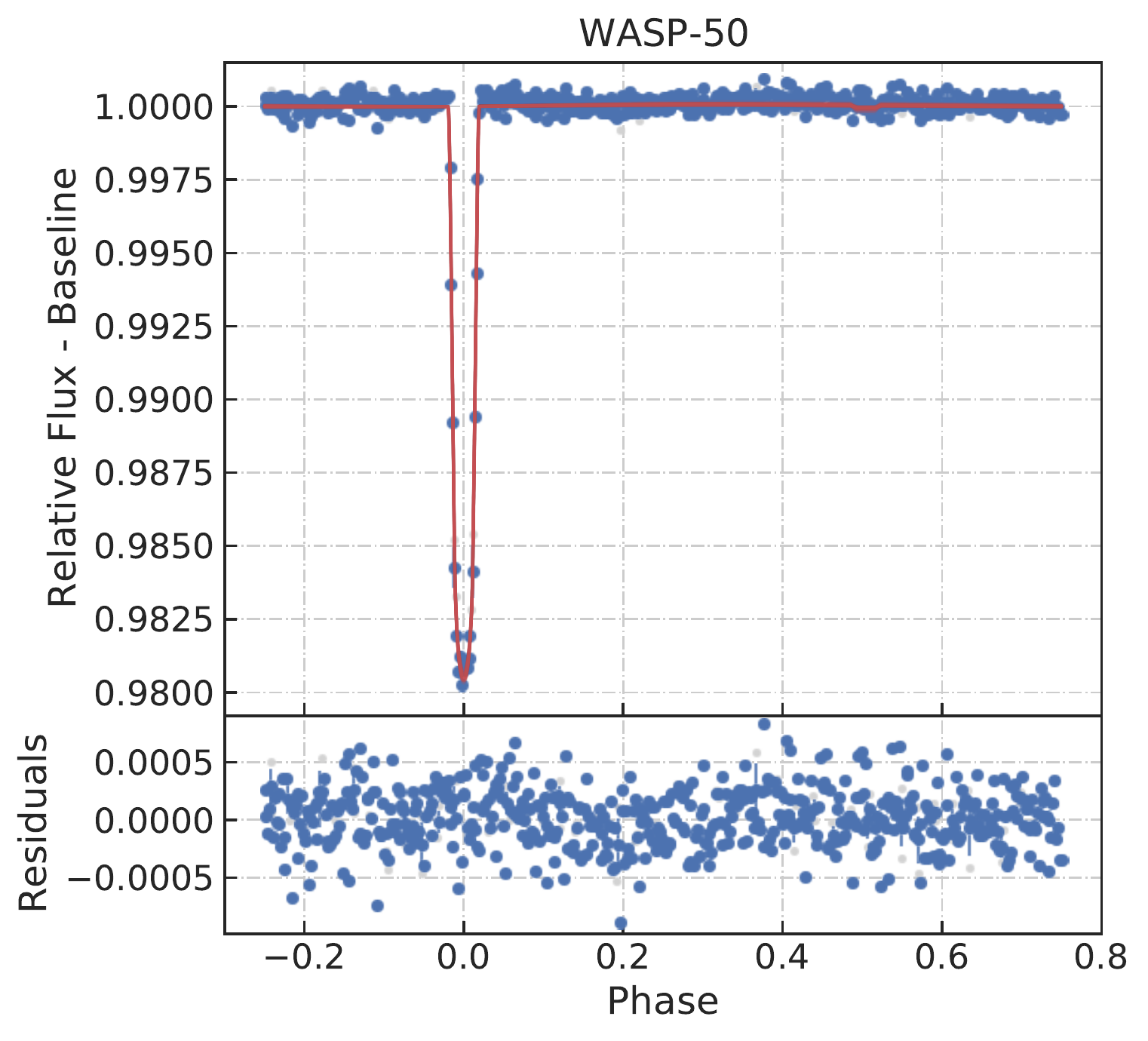}\\
	\includegraphics[height=7.8cm]{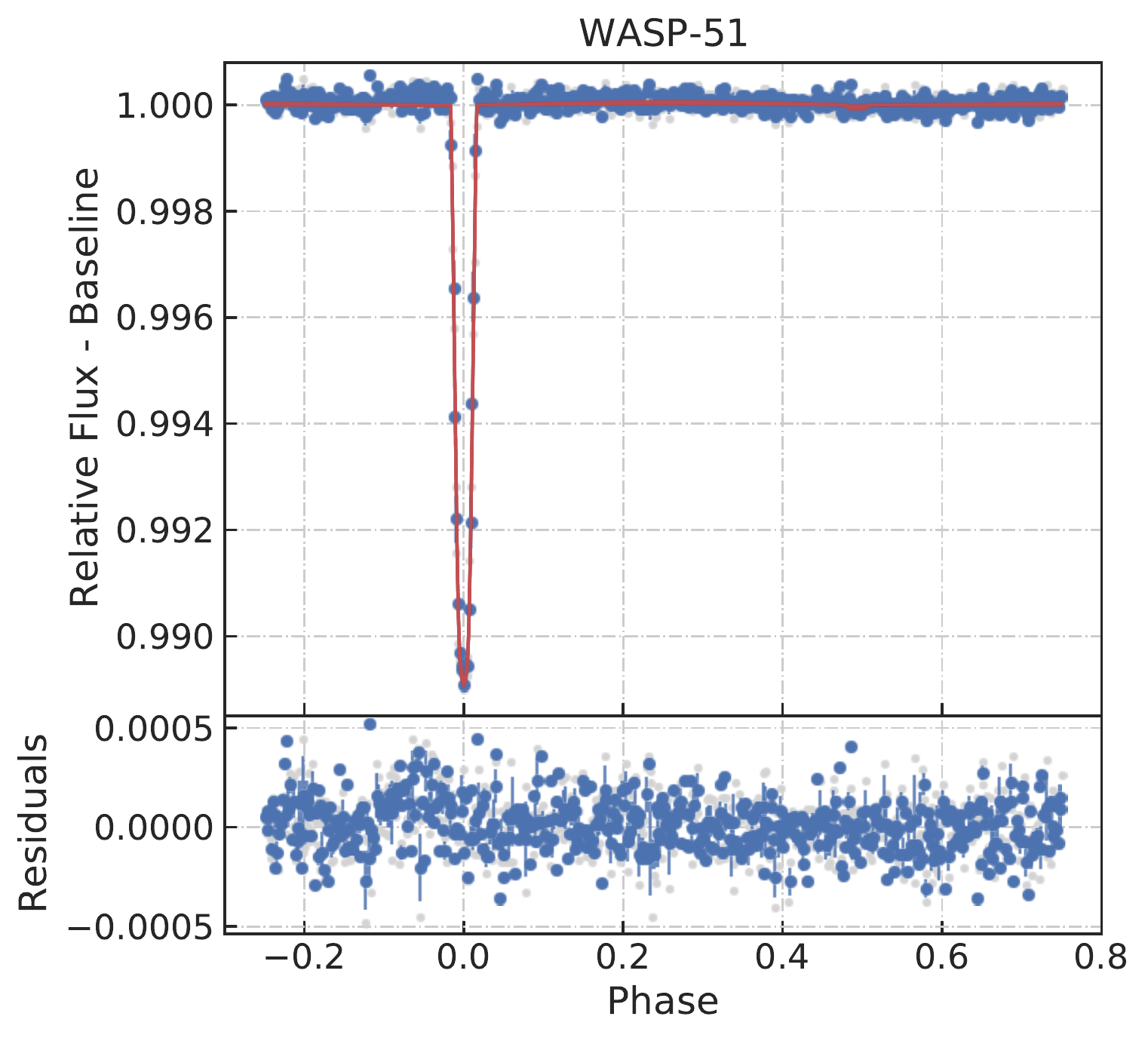}
    \caption{\emph{TESS} phase curves {f}itted by \texttt{allesfitter} software package. The light grey points are the original data binned per 5 minutes, the blue points with error bars (mostly not visible) are the data binned per 15 minutes, and the red curves (seen as one) show 50 models which are randomly drawn from the posteriors.}
	\label{fig:allesPhaseCurves}
\end{figure*}

\begin{figure*}
	\includegraphics[height=7.8cm]{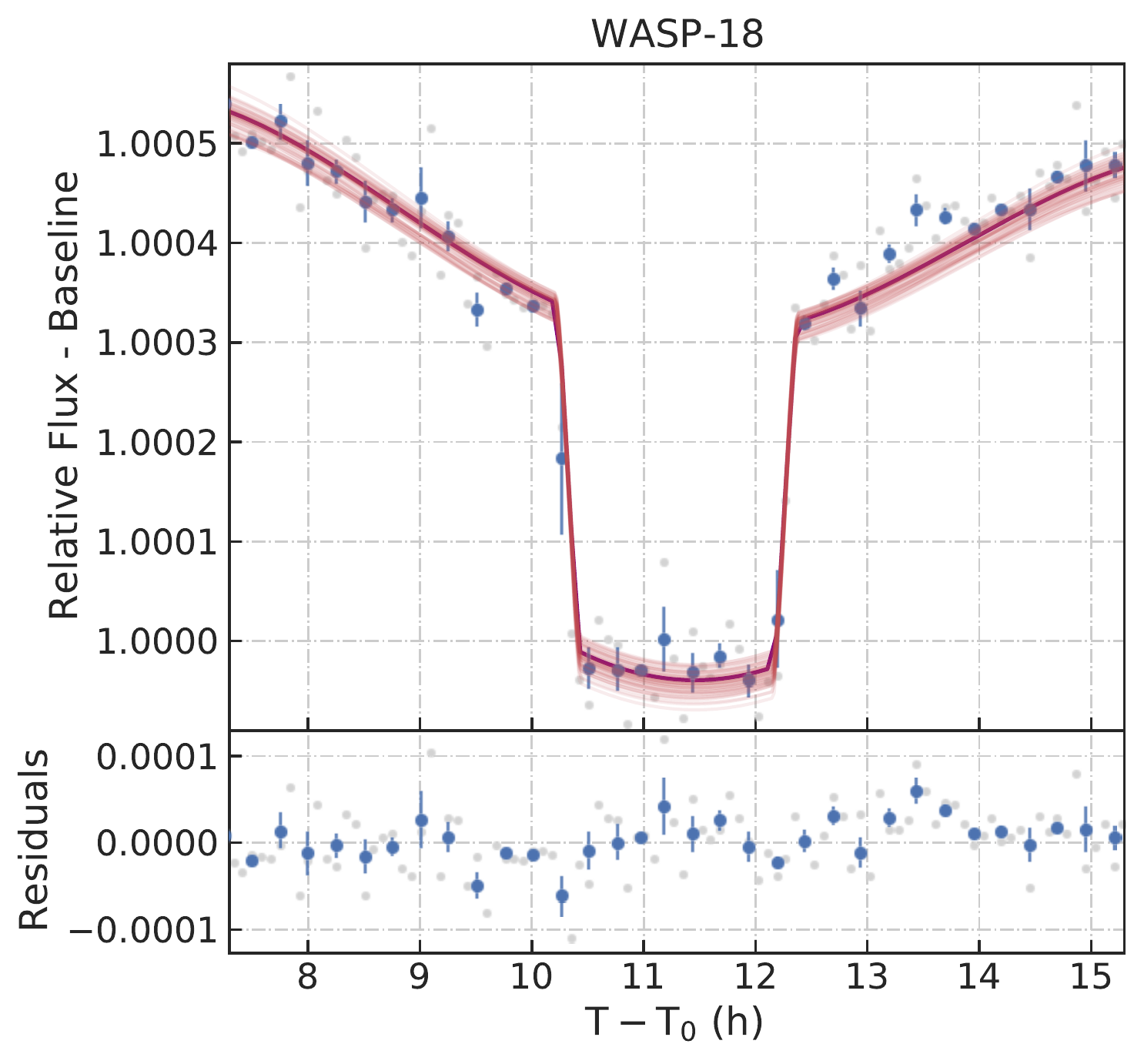}
	\includegraphics[height=7.8cm]{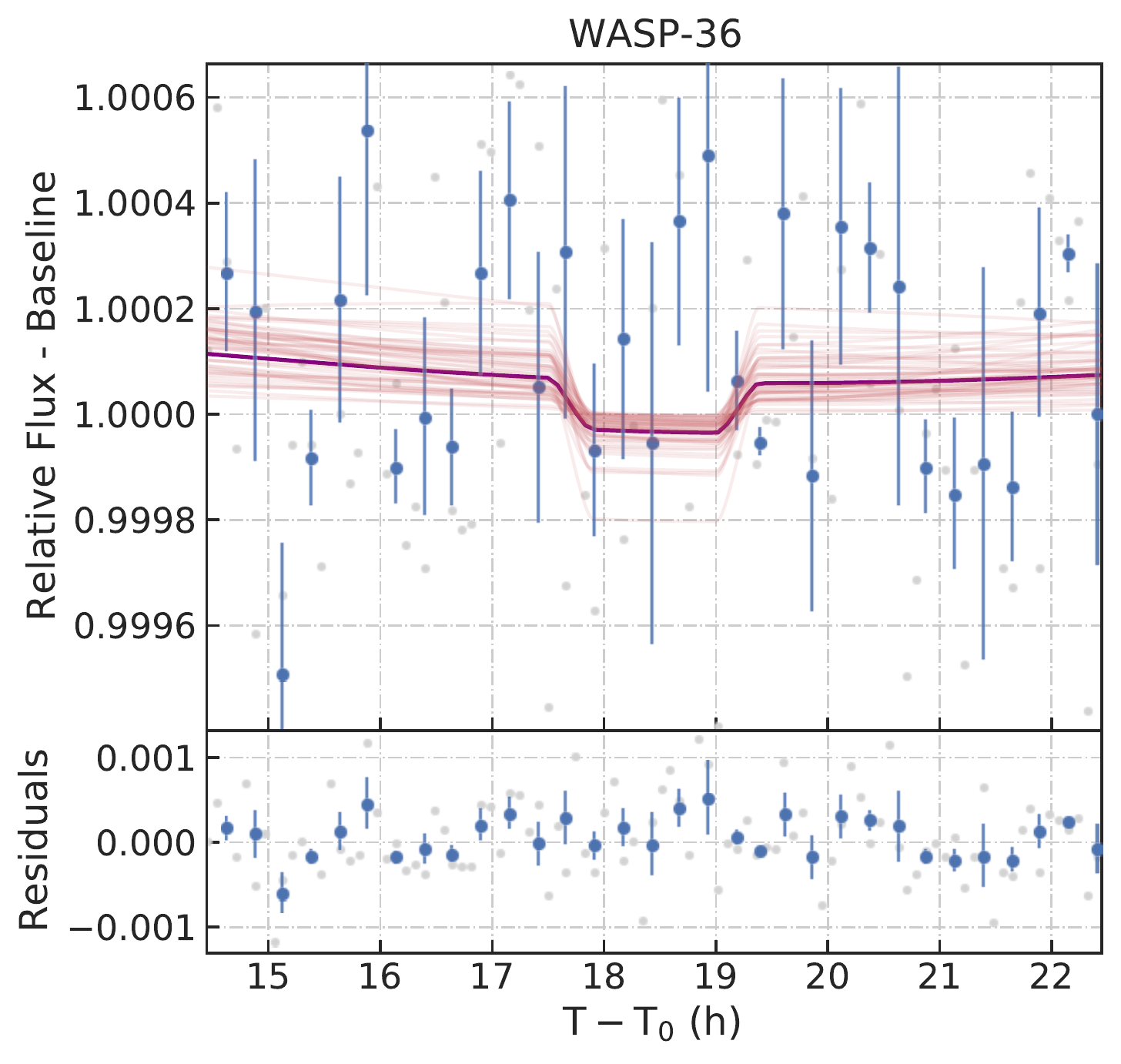}\\
	\includegraphics[height=7.8cm]{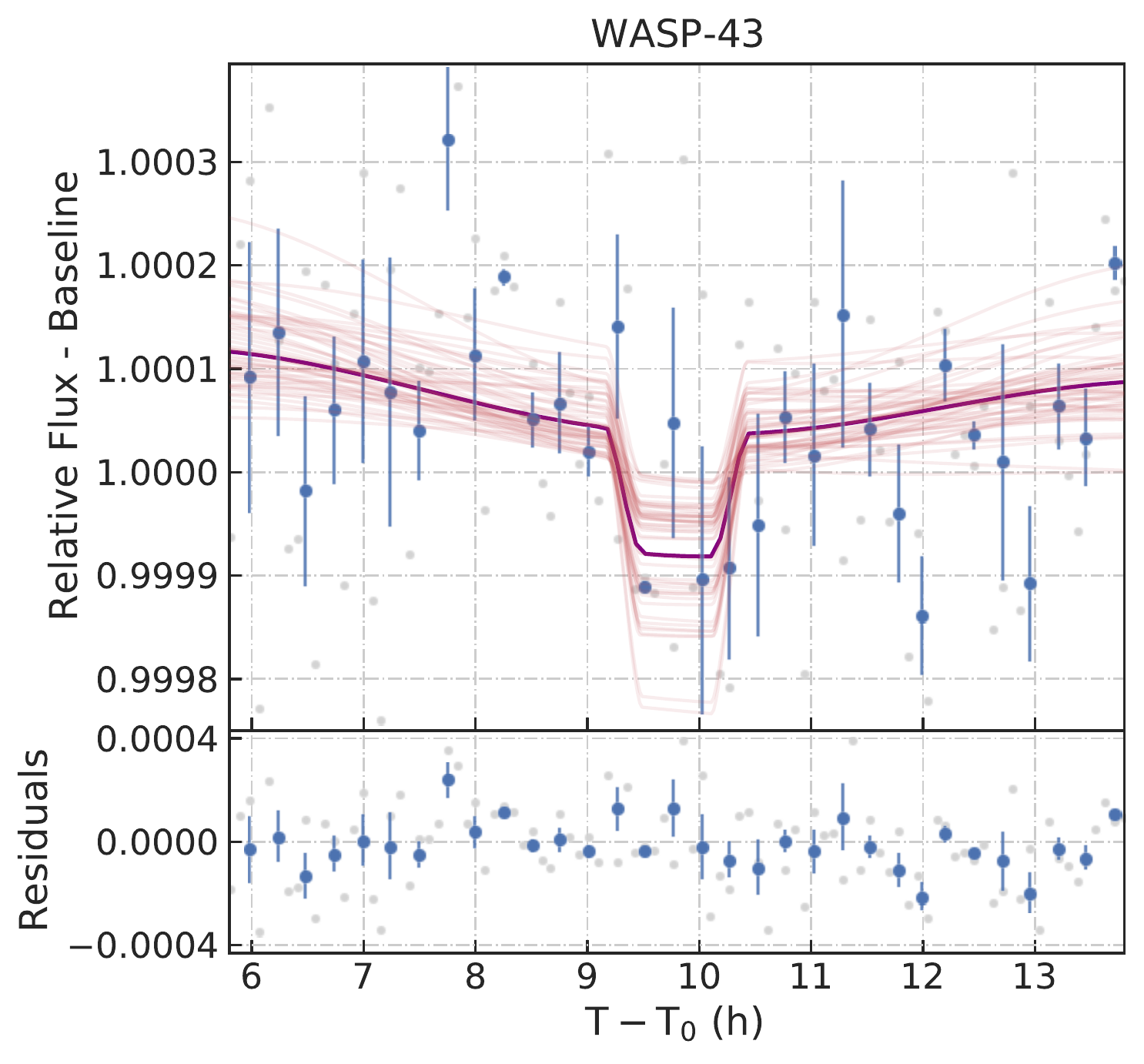}
	\includegraphics[height=7.8cm]{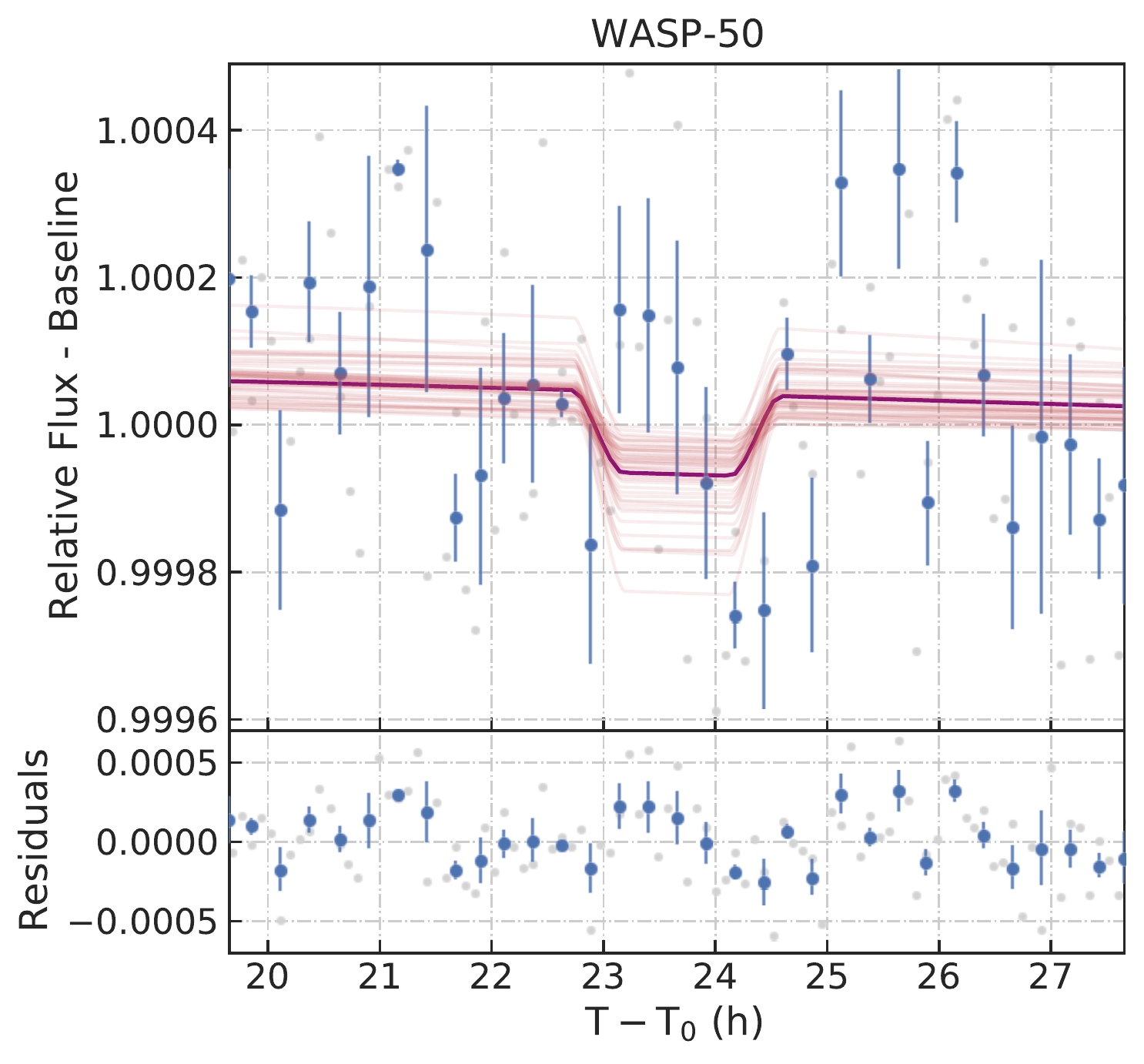}\\
	\includegraphics[height=7.8cm]{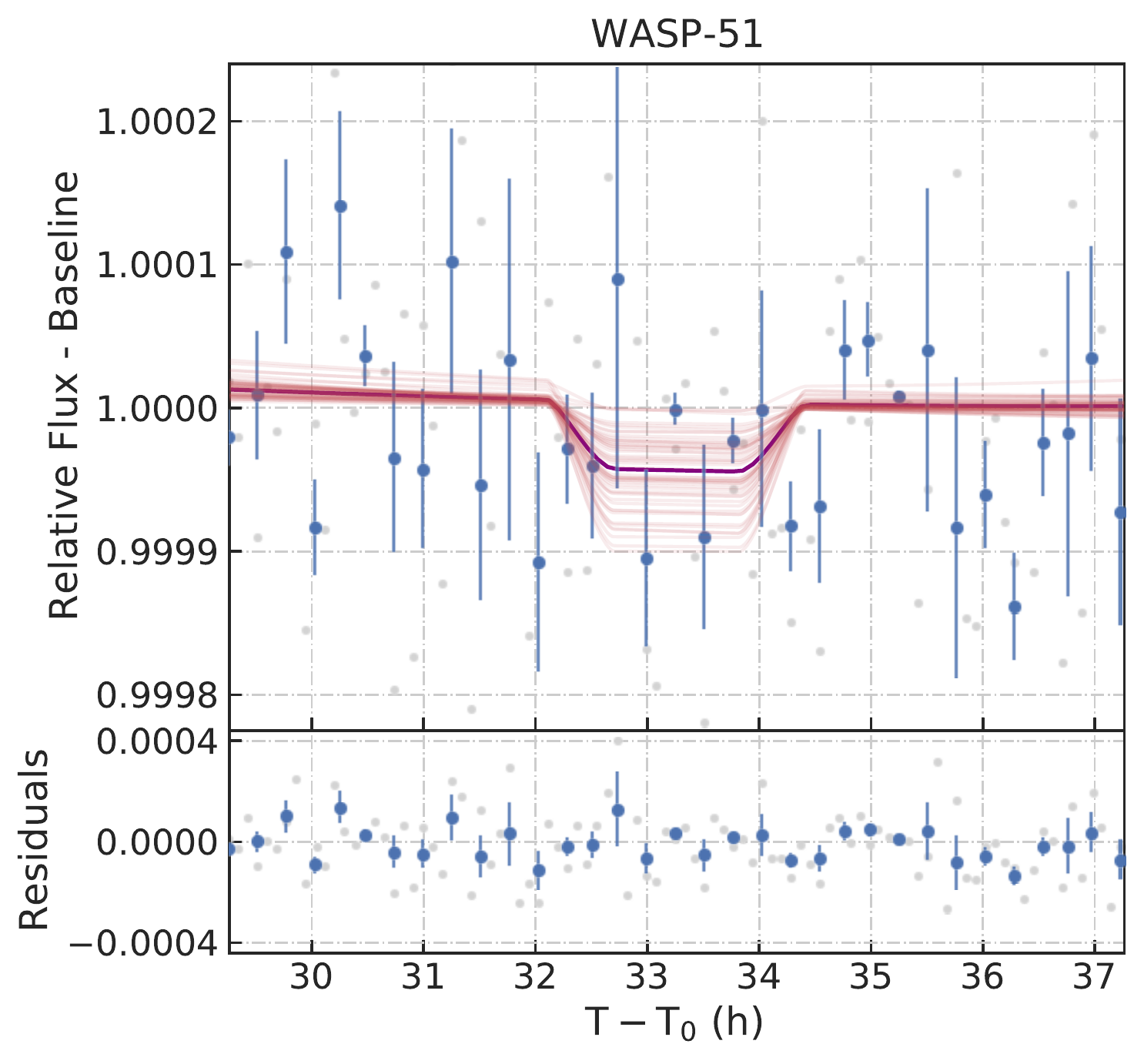}
	\caption{\emph{TESS} phase curves {f}itted by \texttt{allesfitter} software package with the occultation part zoomed. The purple curves show the median model drawn from the posteriors. The meaning of all the points and the red curves is the same as in Fig.~\ref{fig:allesPhaseCurves}.}
	\label{fig:allesOccultations}
\end{figure*}

In what follows, we describe our results from the \emph{TESS} data for each planet in turn. The estimated values of the {f}itted parameters are shown in Table~\ref{tab:allesFitted}, alongside the {f}ixed parameters. In Table~\ref{tab:allesDerrived}, we summarise the parameters subsequently derived from the best-{f}itting phase curve parameters. Fig.~\ref{fig:allesPhaseCurves} and \ref{fig:allesOccultations} show the {f}itted phase curves and occultations, respectively.

\subsubsection{WASP-18}
The \textit{TESS} phase curve of this system has previously been studied by \citet{shporer}, as well as \citet{guenther2021} who also used \texttt{allesfitter} to {f}it the phase curve. We detected a primary transit depth of $10.617_{-0.015}^{+0.014}$\,ppt and an occultation depth of $0.345\pm0.011$\,ppt. The occultation depth is consistent with the values derived by both \citet{shporer} and \citet{guenther2021}, while the primary transit depth we derive is consistent with that of \citet{guenther2021}. \citet{shporer} obtain a transit depth of 9.439$^{+0.027}_{-0.026}$\,ppt; the discrepancy between our value and theirs may be due to di{f}ferent analysis methods and the fact that we used data from four \emph{TESS} sectors, while only two sectors were available at the time of their study. We determined the amplitude of the atmospheric contribution to the phase-curve model to be $0.3065\pm0.0086$\,ppt (i.e., a semi-amplitude of $0.1533\pm0.0043$\,ppt), which lies between the values derived by \citet{guenther2021} and \citet{shporer}. Furthermore, our value is consistent with that of \citet{guenther2021} to within $2\sigma$, which is expected since \texttt{allesfitter} was used for both analyses.

We further estimate the optical albedo of WASP-18\,b based on the measured occultation depth. Due to the high dayside temperature of WASP-18\,b, its thermal emission represents a non-negligible contribution in the \textit{TESS} band, unlike the cooler targets in our sample. The way in which this thermal contribution is estimated may therefore have a signi{f}icant e{f}fect on the resulting albedo constraint. Using Equation \ref{eq:deltaOccTh}, the thermal contribution to the occultation depth in the \textit{TESS} band is 97\,ppm, resulting in an albedo value of $A_\mathrm{g}=0.340\pm0.044$. This albedo calculation assumes e{f}{f}icient day-night energy redistribution ($f=0.25$) in the estimation of $T_\mathrm{p}$. However, existing infrared observations of WASP-18\,b indicate that its day-night energy redistribution is ine{f}{f}icient (e.g., \citealt{Arcangeli2018}, see also Section~\ref{sec:atmospheres}). 

A more accurate albedo constraint can be derived by considering a more realistic thermal contribution to the \textit{TESS} occultation depth. \citet{shporer} used the atmospheric model of \citet{Arcangeli2018}, found by {f}itting the \textit{HST} and \textit{Spitzer} occultation depths of WASP-18\,b and resulting in a thermal contribution of 0.327\,ppt. As noted by \citet{shporer}, this contribution is consistent with the observed occultation depth, meaning that only an upper limit can be placed on the re{f}lected contribution. They placed a $2\sigma$ upper limit of $A_\mathrm{g}<0.048$. To derive the geometric albedo, we used our value of the detected occultation depth and their value of the thermal contribution of 0.327\,ppt, and we come to $A_\mathrm{g}=0.025\pm0.027$ which is consistent with their value obtained from the upper limit on the occultation depth. The high values of the uncertainties are caused by uncertainties of the thermal contribution which are expected to be a few per cent using the model of \citealt{Arcangeli2018} as in \citet{shporer}. Thus, we set them to be 5\,per cent when calculating the geometric albedo uncertainties. However, as our detected occultation depth is very similar to theirs (0.345 vs 0.341\,ppt), we can also not claim a detection of the re{f}lected light since the di{f}ference between the thermal emission and our occultation depth is not at $3\sigma$ signi{f}icance ($\sim1.6\sigma$). We, therefore, set a $3\sigma$ uppper limit on the re{f}lected light by the planet of $<0.033$\,ppt implying an upper limit on the geometric albedo $A_\mathrm{g}<0.045$, consistent with the $2\sigma$ upper limit of \citealt{shporer}.

We note that our self-consistent atmospheric model for WASP-18\,b, discussed in Section~\ref{sec:atmospheres}, is consistent with the observed occultation depth within 2$\sigma$ without the inclusion of scattering from clouds or hazes. The predicted thermal contribution from this model is slightly higher than the observed occultation depth, and is therefore consistent with zero albedo in the \textit{TESS} band.

\subsubsection{WASP-36}
We detected the primary transit of WASP-36\,b and obtained a transit depth of $18.67\pm0.13$\,ppt. This value is lower than that obtained by \citet{maciejewski} in the $R$-band ($19.349\pm0.320$\,ppt), however this di{f}ference may be due to the di{f}ferent wavelength range used. We obtained an occultation depth of $0.105^{+0.057}_{-0.049}$\,ppt, and therefore did not signi{f}icantly detect the occultation of WASP-36\,b ($\sim2\sigma$ detection). This is a result of the relatively high RMS$_\mathrm{w}$ of the data of 0.288\,ppt. We place a 3$\sigma$ upper limit on the occultation depth of $\delta_\mathrm{occ}<0.159$\,ppt. This is consistent with the constraint from \citet{wong}, who derive $\delta_\mathrm{occ}=0.09^{+0.10}_{-0.07}$\,ppt using \textit{TESS}. \citet{zhou} derive an occultation depth of $1.3\pm0.4$\,ppt in the $Ks$-band; for the shorter wavelengths at which \textit{TESS} operates, the occultation depth is indeed expected to be lower under the assumption of little or no optical scattering. The upper limit which we derive on the occultation depth corresponds to a $3\sigma$ upper limit on the geometric albedo of $A_\mathrm{g}<0.286$. This is consistent with the geometric albedo constraint derived by \citet{wong} ($A_\mathrm{g}=0.16^{+0.16}_{-0.15}$). 

\subsubsection{WASP-43}
We detected the primary transit of WASP-43\,b, obtaining a transit depth of $26.597^{+0.078}_{-0.070}$\,ppt. This value is slightly di{f}ferent than $25.415\pm0.131$\,ppt in the optical band ($i'+g'$ {f}ilters) published in \citet{hoyer2016}. We do not detect the occultation of WASP-43\,b at su{f}{f}iciently high signi{f}icance, obtaining $\delta_\mathrm{occ}=0.123^{+0.059}_{-0.048}$\,ppt ($2.3\sigma$) while the RMS$_\mathrm{w}$ of our data is 0.148\,ppt. This is consistent with the results of \citet{wong}, who obtain a \textit{TESS} occultation depth of $\delta_\mathrm{occ}=0.17\pm0.07$\,ppt. Furthermore, \citet{chen} measured the occultation depth of WASP-43\,b in the $i'$ band (centred on roughly the same wavelength as the \textit{TESS} bandpass) to be $\delta_\mathrm{occ}=0.37\pm0.22$\,ppt, which is also consistent with our value. \citet{fraine2021} put for the re{f}lected light component a $3\sigma$ upper limit $\delta_\mathrm{occ}<0.067$\,ppt from \textit{HST} WFC3/UVIS data and from this value they derived a $3\sigma$ upper limit $A_\mathrm{g}\lesssim0.06$. From our $3\sigma$ upper limit on the occultation depth of $\delta_\mathrm{occ}<0.161$\,ppt, we derive an upper limit on the geometric albedo of $A_\mathrm{g}<0.154$. This value is consistent with the albedos derived by \citet{wong} and \citet{chen}, i.e., $0.12\pm0.06$ and $0.31\pm0.22$, respectively. Our $\delta_\mathrm{occ}$ and $A_\mathrm{g}$ upper limits are also consistent with the upper limits obtained by \citet{fraine2021}.

\subsubsection{WASP-50}
We obtained a transit depth value of $19.502^{+0.085}_{-0.093}$\,ppt. This is consistent with the derived value of $19.321\pm0.167$\,ppt in the $I$ and $R$ bands published in \citet{chakrabarty}. We detected an occultation with less than $3\sigma$ signi{f}icance, $\delta_\mathrm{occ}=0.117^{+0.051}_{-0.048}$\,ppt ($\sim2.4\sigma$), which is nevertheless the {f}irst occultation measurement of this system. Our derived value is lower than the expected value of 0.342\,ppt (assuming $A_g=1$) and also lower than RMS$_\mathrm{w}$ of our data, 0.174\,ppt. We placed a $3\sigma$ upper limit on the occultation depth $\delta_\mathrm{occ}<0.149$\,ppt. From the upper limit of the occultation depth, we then derived an upper limit on the geometric albedo of $A_\mathrm{g}<0.44$. 

\subsubsection{WASP-51}
We obtained a transit depth of $10.872\pm0.055$\,ppt. While this value is not consistent with the transit depths derived by \citet{maciejewski} and \citet{saha} in the $R$ and $V$ bands, respectively, the di{f}ference may be due to the use of di{f}ferent wavelength ranges. Indeed, \citet{saeed} discovered a strong dependency of the transit depth with wavelength. The occultation of WASP-51\,b was not detected at su{f}{f}iciently high signi{f}icance, with a measured occultation depth of $0.048^{+0.033}_{-0.024}$\,ppt ($\sim1.7\sigma$). As in the case of WASP-50, this the {f}irst occultation measurement of this system. The RMS of the data was also high, at 0.105\,ppt. We placed an upper limit on the occultation depth $\delta_\mathrm{occ}<0.086$\,ppt. This allowed us to set a $3\sigma$ upper limit on the geometric albedo $A_\mathrm{g}<0.368$\,ppt.

\begin{table*}
	\centering
	\caption{Results of the analysed data sets. Parameters presented and their meaning is following: the used HAWK-I {f}ilter, RMS$_\mathrm{w}$ is the standard deviation of the weighted mean of the binned data sets, $\delta_\mathrm{occ,re,exp}$ is the expected occultation depth due to re{f}lected light, $\delta_\mathrm{occ,th,exp}$ is the expected occultation depth due to thermal radiation, $\delta_\mathrm{occ,exp}$ is the total expected occultation depth (the sum of the both previous), $\delta_\mathrm{occ}$ is the derived occultation depth (with at least $3\sigma$ signi{f}icance), $\delta_{\mathrm{occ,}3\sigma\mathrm{UL}}$ is the $3\sigma$ upper limit on the occultation depth, and $A_\mathrm{g}$ is the upper limit (or derived value) of the geometric albedo. For \textit{TESS} data sets $\delta_\mathrm{tra}$ is the inferred transit depth. In the third table $T_\mathrm{eq,p}$ is the equilibrium temperature of the planet. \emph{Notes}: $^{(a)}$:~calculated from Equation~\ref{eq:deltaOccRe} supposing $A_\mathrm{g}=1$; $^{(b)}$:~calculated from Equation~\ref{eq:deltaOccTh} substituting $T_\mathrm{p}=T_\mathrm{eq,p}$ and $T_\star=T_\mathrm{e{f}f,\star}$; $^{(c)}$:~adopted from \citet{Arcangeli2018}, as in \citet{shporer}; $^{(d)}$:~derived using the thermal contribution from \citet{Arcangeli2018}, as in \citet{shporer}; $^{(e)}$:~calculated from Equation~\ref{eq:teq} assuming $f=1/4$ and $A_\mathrm{B}=0$.}
	\label{tab:upper_limits}
	\begin{tabular}{lccccccccl}
	    \hline
	   		\multicolumn{9}{c}{\bf \textit{TESS} data sets} \\
		\hline
	 	System & $\delta_\mathrm{tra}$ & RMS$_\mathrm{w}$ & $\delta_\mathrm{occ,re,exp}$ $^{(a)}$ & $\delta_\mathrm{occ,th,exp}$ $^{(b)}$ & $\delta_\mathrm{occ,exp}$ & $\delta_\mathrm{occ}$ & $A_\mathrm{g}$ & $\delta_{\mathrm{occ,3\sigma UL}}$ & $A_\mathrm{g,3\sigma UL}$\\
	 	 & [ppt] & [ppt] & [ppt] & [ppt] & [ppt] & [ppt] & & [ppt] & \\
	    \hline
	    	WASP-18 & $10.617_{-0.015}^{+0.014}$ & 0.042 & 0.690 & ~~~~~0.327$^{(c)}$ & 1.017 & $0.345_{-0.011}^{+0.011}$ & ~~~~~~$0.025^{+0.027~(d)}_{-0.027}$ & ~-- & $<0.045^{(d)}$ \\
		    WASP-36 & $18.670^{+0.130}_{-0.130}$ & 0.288 & 0.500 & 0.012 & 0.512 & $0.105_{-0.049}^{+0.057}$ & $0.181^{+0.112}_{-0.096}$ & $<0.159$ & 
		    $<0.286$ \\
		    WASP-43 & $26.597_{-0.070}^{+0.078}$ & 0.148  & 0.990 & 0.007 & 0.997 & $0.123_{-0.048}^{+0.059}$ & $0.116^{+0.059}_{-0.048}$ & $<0.161$ & $<0.154$ \\
		    WASP-50 & $19.502_{-0.093}^{+0.085}$ & 0.174 & 0.340 & 0.002 & 0.342 & $0.117_{-0.048}^{+0.051}$ & $0.344^{+0.156}_{-0.148}$ & $<0.149$ & $<0.440$ \\
		    WASP-51 & $10.872^{+0.055}_{-0.055}$ & 0.105 & 0.263 & 0.004 & 0.267 & $0.048_{-0.024}^{+0.033}$ & $0.197^{+0.149}_{-0.109}$ & $<0.086$ & $<0.368$ \\
		\hline
	\end{tabular}
	\begin{tabular}{lcccccrcc}
		\hline	
			\multicolumn{7}{c}{\bf VLT HAWK-I data set} \\
		\hline
		    System & Filter & RMS$_\mathrm{w}$ [ppt] & $\delta_\mathrm{occ,re,exp}$ [ppt]$^{(a)}$ & $\delta_\mathrm{occ,th,exp}$ [ppt]$^{(b)}$ & $\delta_\mathrm{occ,exp}$ [ppt] & $\delta_\mathrm{occ}$ [ppt] \\
		\hline
			WASP-43 & NB2090 & 0.298 & 0.990 & 0.750 & 1.740 & $1.26^{+0.17}_{-0.17}$ \\
	    \hline
	\end{tabular}
	\begin{tabular}{cccccc}
		\hline
		Planet: & WASP-18\,b & WASP-36\,b & WASP-43\,b & WASP-50\,b & WASP-51\,b \\
		\hline
		$T_\mathrm{eq}$ [K]$^{(e)}$ & $2504^{+63}_{-65}$ & $1724^{+43}_{-43}$ & $1439^{+34}_{-31}$ & $1393^{+42}_{-42}$ & $1637^{+42}_{-42}$ \\
		\hline
	\end{tabular}
\end{table*}

\movetabledown=1.33in
	\begin{longrotatetable}
	    \tablecaption{Posterior values of all the {f}itted parameters (and hyperparameters) of \textit{TESS} phase curves of all the systems analysed in this work obtained by using by \texttt{allesfitter} NS.}
		\label{tab:allesFitted}
		\begin{tabular}{lcccccc}
		\hline
			Parameter~/~System & WASP-18 & WASP-36 & WASP-43 & WASP-50 & WASP-51 & {f}it/{f}ixed \\ 
        \hline 
        $R_\mathrm{p}/R_\star$ & $0.09669\pm0.00013$ & $0.13271\pm0.00085$ & $0.15865\pm0.00044$ & $0.13668\pm0.00058$ & $0.10949\pm0.00046$ & {f}it \\ 
        $(R_\star + R_\mathrm{p})/a$ & $0.3131\pm0.0023$ & $0.1952\pm0.0044$ & $0.2474\pm0.0022$ & $0.1550\pm0.0022$ & $0.1609\pm0.0013$ & {f}it \\ 
        $\cos{i}$ & $0.0997_{-0.0063}^{+0.0059}$ & $0.1157\pm0.0059$ & $0.1481\pm0.0027$ & $0.0974\pm0.0028$ & $0.1214\pm0.0013$ & {f}it \\ 
        $T_{0}$ [epoch] & $0.000021\pm0.000035$ & $-0.00002\pm0.00014$ & $0.000013\pm0.000048$ & $-0.000017\pm0.000096$ & $0.00004\pm0.00012$ & {f}it \\ 
        $P$ [d] & $0.9414518$ & $1.5373653$ & $0.81347753$ & $1.9550959$ & $2.8106084$ & {f}ixed \\ 
        $\sqrt{e} \cos{\omega}$ & $-0.00163718$ & $0.0$ & $0.05017120$ & $0.06824257$ & $-0.05781180$ & {f}ixed \\ 
        $\sqrt{e} \sin{\omega}$ & $-0.09379403$ & $0.0$ & $-0.03135045$ & $0.06590108$ & $-0.17792638$ & {f}ixed \\ 
        $J_\mathrm{TESS}$ & $0.0041\pm0.0012$ & $0.0019_{-0.0013}^{+0.0024}$ & $0.0032_{-0.0018}^{+0.0022}$ & $0.0035_{-0.0021}^{+0.0029}$ & $0.0037_{-0.0021}^{+0.0026}$ & {f}it \\ 
        $q_{1;~\mathrm{TESS}}$ & $0.36660057$ & $0.38667813$ & $0.53140888$ & $0.44026143$ & $0.37665187$ & {f}ixed \\ 
        $q_{2;~\mathrm{TESS}}$ & $0.15570471$ & $0.15819508$ & $0.10952400$ & $0.13763435$ & $0.1549222$ & {f}ixed \\ 
        $\ln\sigma_\mathrm{TESS}$ [ln(relative {f}lux)] & $-10.075\pm0.043$ & $-7.774\pm0.034$ & $-8.605\pm0.048$ & $-8.245\pm0.030$ & $-8.732\pm0.026$ & {f}it \\ 
        $\Delta F_\mathrm{TESS}$ & $-0.0002474\pm0.0000076$ & $0.000141_{-0.000040}^{+0.000038}$ & $0.000161_{-0.000027}^{+0.000025}$ & $0.000050_{-0.000019}^{+0.000017}$ & $0.0000334\pm0.0000099$ & {f}it \\ 
        $A_\mathrm{p;~beaming;~TESS}$ [ppt] (semi-amplitude) & $0.0312\pm0.0035$ & $0.032_{-0.018}^{+0.024}$ & $0.0157_{-0.0096}^{+0.013}$ & $0.033_{-0.014}^{+0.015}$ & $0.0139_{-0.0069}^{+0.0074}$ & {f}it \\ 
        $A_\mathrm{p;~atmospheric;~TESS}$ [ppt] (amplitude) & $0.3065\pm0.0086$ & $0.062_{-0.036}^{+0.049}$ & $0.037_{-0.022}^{+0.030}$ & $0.044_{-0.024}^{+0.030}$ & $0.0031_{-0.0022}^{+0.0042}$ & {f}it \\
        $A_\mathrm{p;~ellipsoidal;~TESS}$ [ppt] (amplitude) & $0.3516\pm0.0089$ & $0.099_{-0.050}^{+0.058}$ & $0.085\pm0.037$ & $0.0084_{-0.0057}^{+0.0110}$ & $0.026_{-0.014}^{+0.016}$ & {f}it \\ 
		\hline
		\end{tabular}
	\end{longrotatetable}
\movetabledown=1.3in
	\begin{longrotatetable}
		\tablecaption{Posterior values of all the derived parameters of \emph{TESS} phase curves of all the systems analysed in this work obtained by using by \texttt{allesfitter} NS.}
		\label{tab:allesDerrived}
			\begin{tabular}{lccccc}
		\hline
            Parameter~/~System & WASP-18 & WASP-36 & WASP-43 & WASP-50 & WASP-51 \\ 
		\hline 
        Host star radius over semi-major axis; $R_\star/a$ & $0.2855\pm0.0021$ & $0.1723\pm0.0038$ & $0.2135\pm0.0018$ & $0.1364\pm0.0019$ & $0.1450\pm0.0012$ \\ 
        Semi-major axis over host star radius; $a/R_\star$ & $3.502\pm0.025$ & $5.80_{-0.12}^{+0.13}$ & $4.683\pm0.041$ & $7.33\pm0.11$ & $6.895\pm0.056$ \\ 
        Planetary radius over semi-major axis; $R_\mathrm{p}/a$ & $0.02761\pm0.00023$ & $0.02286\pm0.00062$ & $0.03388\pm0.00037$ & $0.01864\pm0.00033$ & $0.01588\pm0.00018$ \\ 
        Planetary radius; $R_\mathrm{p}$ [$\mathrm{R_{\oplus}}$] & $13.29\pm0.42$ & $13.6\pm2.5$ & $11.54\pm0.18$ & $13.1\pm1.6$ & $15.89\pm0.37$ \\ 
        Planetary radius; $R_\mathrm{p}$ [$\mathrm{R_{J}}$] & $1.186\pm0.038$ & $1.21\pm0.22$ & $1.030\pm0.016$ & $1.17\pm0.15$ & $1.417\pm0.033$ \\ 
        Semi-major axis of the planetary orbit; $a$ [$\mathrm{R_{\odot}}$] & $4.41\pm0.14$ & $5.46\pm1.0$ & $3.124\pm0.054$ & $6.45\pm0.81$ & $9.17\pm0.22$ \\ 
        Semi-major axis of the planetary orbit; $a$ [\textsc{au}] & $0.02052\pm0.00067$ & $0.0254\pm0.0047$ & $0.01453\pm0.00025$ & $0.0300\pm0.0038$ & $0.0427\pm0.0010$ \\
        Inclination angle of the planetary orbit; $i$ [deg] & $84.28_{-0.34}^{+0.37}$ & $83.36\pm0.34$ & $81.49\pm0.16$ & $84.41\pm0.16$ & $83.026\pm0.077$ \\ 
        Impact parameter; $b_\mathrm{tra}$ & $0.352_{-0.020}^{+0.018}$ & $0.671_{-0.020}^{+0.018}$ & $0.6946\pm0.0069$ & $0.7099_{-0.0110}^{+0.0099}$ & $0.8651\pm0.0029$  \\ 
        Total transit duration; $T_\mathrm{tot}$ [h] & $2.1956\pm0.0041$ & $1.867\pm0.017$ & $1.2552\pm0.0058$ & $1.810\pm0.012$ & $2.257\pm0.011$ \\
        Full-transit duration; $T_\mathrm{full}$ [h] & $1.7486\pm0.0043$ & $1.121_{-0.027}^{+0.025}$ & $0.6393\pm0.0091$ & $1.000_{-0.019}^{+0.018}$ & $0.686_{-0.039}^{+0.037}$ \\
        Epoch occultation; $T_\mathrm{0;occ}$ & $0.470655\pm0.000035$ & $0.76867\pm0.00014$ & $0.408289\pm0.000049$ & $0.985589\pm0.000096$ & $1.38599\pm0.00012$ \\ 
        Impact parameter occultation; $b_\mathrm{occ}$ & $0.346_{-0.020}^{+0.018}$ & $0.671_{-0.020}^{+0.018}$ & $0.6921\pm0.0069$ & $0.7188_{-0.011}^{+0.0100}$ & $0.8093\pm0.0027$ \\ 
        Transit depth; $\delta_\mathrm{tra; p; TESS}$ [ppt] & $10.617_{-0.015}^{+0.014}$ & $18.67\pm0.13$ & $26.597_{-0.070}^{+0.078}$ & $19.502_{-0.093}^{+0.085}$ & $10.872\pm0.055$ \\
        Occultation depth; $\delta_\mathrm{occ; p; TESS}$ [ppt] & $0.345\pm0.011$ & $0.105_{-0.049}^{+0.057}$ & $0.123_{-0.048}^{+0.059}$ & $0.117_{-0.048}^{+0.051}$ & $0.048_{-0.024}^{+0.033}$ \\
        Nightside {f}lux of the planet; $F_\mathrm{nightside; p; TESS}$ [ppt] & $0.039\pm0.011$ & $0.034_{-0.023}^{+0.043}$ & $0.079_{-0.043}^{+0.057}$ & $0.068_{-0.041}^{+0.047}$ & $0.045_{-0.025}^{+0.031}$ \\
        Median host star density -- all orbits; $\rho_\mathrm{\star}$ (cgs) & $0.917\pm0.020$ & $1.564_{-0.096}^{+0.11}$ & $2.936\pm0.077$ & $1.950_{-0.080}^{+0.086}$ & $0.785\pm0.019$ \\
		\hline
		\end{tabular}
	\end{longrotatetable}

\subsection{HAWK-I occultation measurement for WASP-43\,b}
\label{sec:upper_limits_hawki}
\begin{figure}
	\includegraphics[width=\columnwidth]{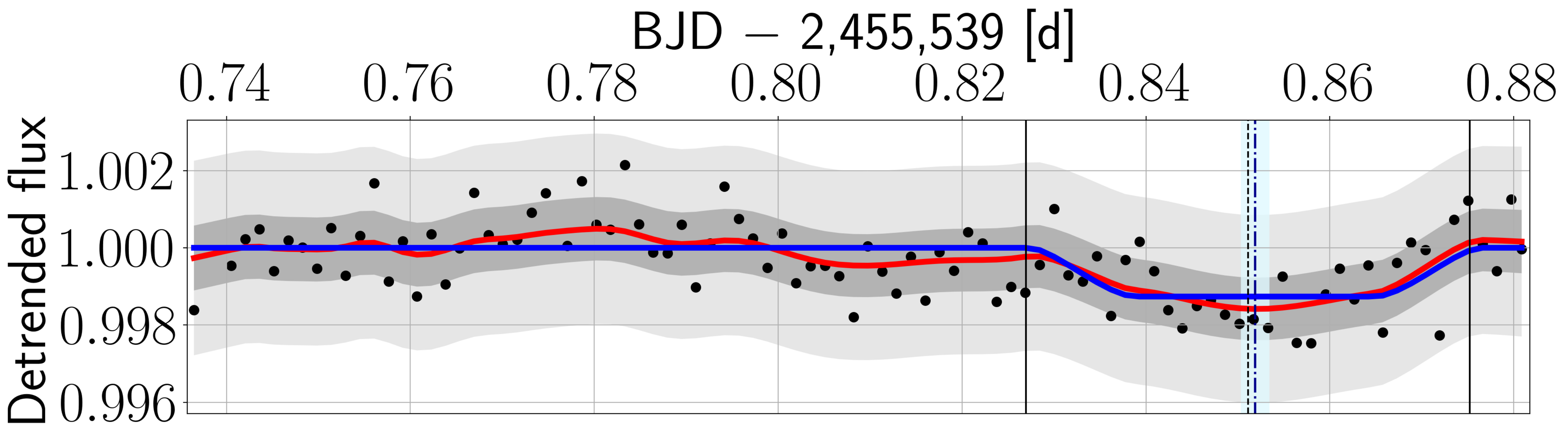} \\
	\includegraphics[width=\columnwidth]{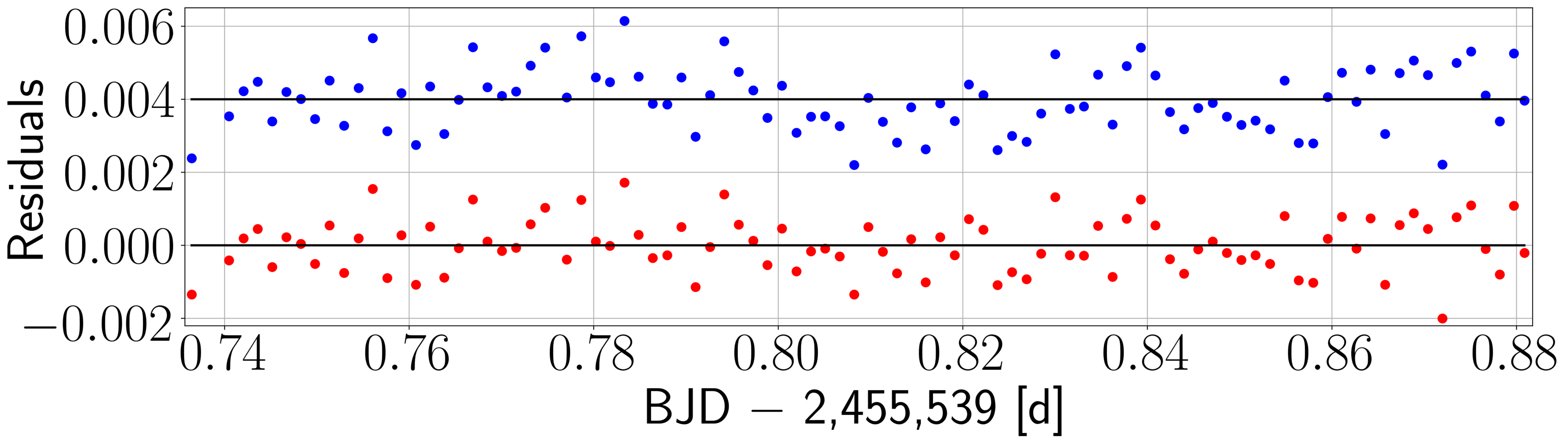} \\	
	    \caption{Results of the best-{f}itting model of WASP-43 from HAWK-I data. \emph{The upper panel}: Detrended occultation light curve. The black dots are our measurements binned per 2 minutes, the red curve shows the {f}it of the data (the occultation model + noise model) together with $1\sigma$ and $3\sigma$ regions depicted as the shaded regions. The blue curve is the occultation model only. The black vertical lines show the calculated beginning, the centre, and the end of the occultation, and the blue dot-and-dash vertical line shows the inferred centre of the occultation together with $1\sigma$ uncertainty region (the light-blue area). \emph{The bottom panel}: blue -- residuals of the occultation model, red -- residuals of the {f}it.}
		\label{fig:w43_fit}
\end{figure}

We detected the occultation of WASP-43\,b using the HAWK-I NB2090 data described in Section \ref{sec:hawki_sets}, consistent with the detection by \citet{gillon2012} who used the same data set but di{f}ferent analysis methods. Here, we {f}irst binned the near-infrared HAWK-I data by 2-minute time intervals. As well as for the \emph{TESS} data sets, we have calculated the $\mathrm{RMS_w}$. The {f}itted light curve is shown in Fig.~\ref{fig:w43_fit}. All used and inferred parameters are summarised in Table~\ref{tab:wasp43_res}. 

We detected the occultation of WASP-43\,b and inferred an occultation depth, $\delta_\mathrm{occ}$, of $1.26\pm0.17$\,ppt. The inferred time of the occultation centre is consistent with the expected value within the derived uncertainty. Our $\delta_\mathrm{occ}$ value is consistent with the value of \citet{gillon2012} ($1.56\pm0.14$\,ppt) within $1.8\sigma$. Our inferred occultation depth of $1.26\pm0.17$\,ppt is signi{f}icantly deeper than our \textit{TESS} occultation depth upper limit of 0.161\,ppt. This implies that planet-star {f}lux ratio is increasing with wavelength, which is naturally explained by the decreasing stellar {f}lux and increasing planetary thermal emission with wavelength in the near-infrared.

We use this occultation depth to calculate the brightness temperature of WASP-43\,b at $\sim2.09\,\upmu$m, obtaining a value of $T_\mathrm{b}=1619\pm52$\,K. This temperature can be used to gain some initial insights into the energy redistribution in the atmosphere of WASP-43\,b. For example, the equilibrium temperature of WASP-43\,b assuming zero albedo and a {f}lux correction factor, $f$, of 1/4 is $T_\mathrm{eq}=1439\pm34$\,K (see Equation~\ref{eq:teq}). The brightness temperature corresponding to the HAWK-I occultation is greater than $T_\mathrm{eq}$, which may be due to ine{f}{f}icient day-night energy redistribution (i.e., $f>1/4$). A lower limit on the e{f}{f}iciency of day-night energy redistribution can be estimated by substituting $T_\mathrm{b}-\sigma_{T_\mathrm{b}}$ for $T_\mathrm{eq,p}$ in Equation~\ref{eq:teq} and solving for $f$. We obtain a physically plausible estimate of $f\ge0.35$, which lies between the limits of $f=1/4$ (uniform redistribution) and $f=2/3$ (instantaneous reradiation). This is consistent with the result obtained by \citet{chen} of $f\ge0.56$, measured in the $K$ band.

While optical observations can be used to estimate the optical albedos of hot Jupiter atmospheres, inferring infrared scattering can be more complex. In the near-infrared, thermal emission dominates the observed planetary {f}lux and is expected to be signi{f}icantly greater than the contribution from re{f}lected light. Furthermore, molecular opacity in the infrared causes the planetary thermal emission to signi{f}icantly deviate from a blackbody spectrum, as can be seen from the evident H$_2$O absorption in the HST/WFC3 spectrum of WASP-43\,b \citep{Kreidberg2014}. As a result, the method described in Section~\ref{sec:upper_limits_tess} to estimate optical geometric albedos should not be used in the near-infrared. Instead, detailed radiative-convective atmospheric models can be used to explain multi-wavelength observations and assess the need for optical and/or infrared scattering. We do this for WASP-43\,b and WASP-18\,b in Section~\ref{sec:atmospheres}, and {f}ind that cloud scattering is not required to explain either of their optical to infrared spectra.

While our self-consistent atmospheric models indicate that cloud scattering is not needed to explain the optical and infrared observations of WASP-43\,b, \citet{keating} {f}ind that an infrared albedo of $0.24\pm0.01$ is needed to {f}it the HST/WFC3 and Spitzer observations. However, we note that their atmospheric model assumes an isothermal temperature pro{f}ile, which does not capture the e{f}fect of molecular absorption features. In contrast to this, we {f}ind that the HST/WFC3 and Spitzer data can be explained by absorption features due to H$_2$O and CO (see Section~\ref{sec:atmospheres}). This highlights the need to consider molecular spectral features when interpreting infrared observations. Nevertheless, in order to compare with the results of \citet{keating}, we use the HAWK-I occultation depth derived above to estimate a nominal infrared albedo. As in \citet{keating}, we assume a blackbody thermal contribution to the observed planetary {f}lux. We use a planetary temperature of 1483\,K, i.e., the best-{f}itting isothermal temperature found by \citet{keating}. Using Equation~\ref{eq:deltaOccTh}, this results in a nominal thermal contribution of 0.864\,ppt. Following the methods outlined in Section~\ref{sec:upper_limits_tess}, this results in an estimated infrared albedo of $A_\mathrm{g}=0.395^{+0.174}_{-0.176}$. Our nominal albedo estimate agrees with the results of \citet{keating} when the same assumptions are made. However, we stress that the infrared thermal contribution should not be assumed to take the form of a blackbody, and that detailed atmospheric models are required to interpret infrared observations. We discuss our self-consistent models in Section \ref{sec:atmospheres}. 

\begin{table}
	\centering
	\caption{Deduced, calculated and {f}ixed parameters of the occultation of WASP-43. Notes: $^{(a)}$:~calculated from Equation~\ref{eq:teq} assuming $f=1/4$ and $A_\mathrm{B}=0$; $^{(b)}$:~calculated from Equation~\ref{eq:tb}; $^{(c)}$:~taken from \citet{hellier}.}
	\label{tab:wasp43_res}
	\begin{tabular}{cr}
		\hline
		\textbf{Deduced parameters}: & \\
		occultation depth $\delta_\mathrm{occ}$ [ppt] & $1.261^{+0.165}_{-0.168}$ \\
		$T_0 - 2{,}450{,}000$ [BJD$_\mathrm{TDB}$] & $0.8519^{+0.0016}_{-0.0015}$ \\
		out-of-occultation {f}lux $f_\mathrm{oot}$ & $1.0027\pm0.0003$ \\
		time gradient $T_\mathrm{grad}$ & $0.0013\pm0.0005$ \\
		time gradient $T^2_\mathrm{grad}$ & $-0.0002\pm0.0002$ \\
		$\xi$ (GP)    & $0.0003877^{+0.0001523}_{-0.0001066}$ \\
		$\eta_t$ (GP) & $0.0081001^{+0.0018084}_{-0.0015249}$ \\
		$\eta_a$ (GP) & $0.0266166^{+0.0018885}_{-0.0017666}$ \\
		$\sigma_\mathrm{w}^2$ (GP) & $0.0007866^{+0.0000878}_{-0.0000703}$ \\
		\hline
		\textbf{Calculated parameters}: & \\
		equilibrium temperature$^{(a)}~T_\mathrm{eq}$ [K] & $1439^{+34}_{-31}$ \\
		brightness temperature$^{(b)}~T_\mathrm{b}$ [K] & $1619^{+47}_{-52}$ \\
		\hline
		\textbf{Fixed parameters}$^{(c)}$: &             \\
		period $P$ [d] & 0.81347404                  \\
		scaled semi-major axis $a/R_\star$ & 5.13    \\
		ratio of the radii $R_\mathrm{p}/R_\star$ & 0.159687    \\
		impact parameter $b$ & 0.66                  \\
		\hline
	\end{tabular}
\end{table}

\section{Atmospheric Constraints for WASP-43\,\lowercase{b} and WASP-18\,\lowercase{b}} \label{sec:atmospheres}

\begin{figure}
    \centering
    \includegraphics[width=0.5\textwidth]{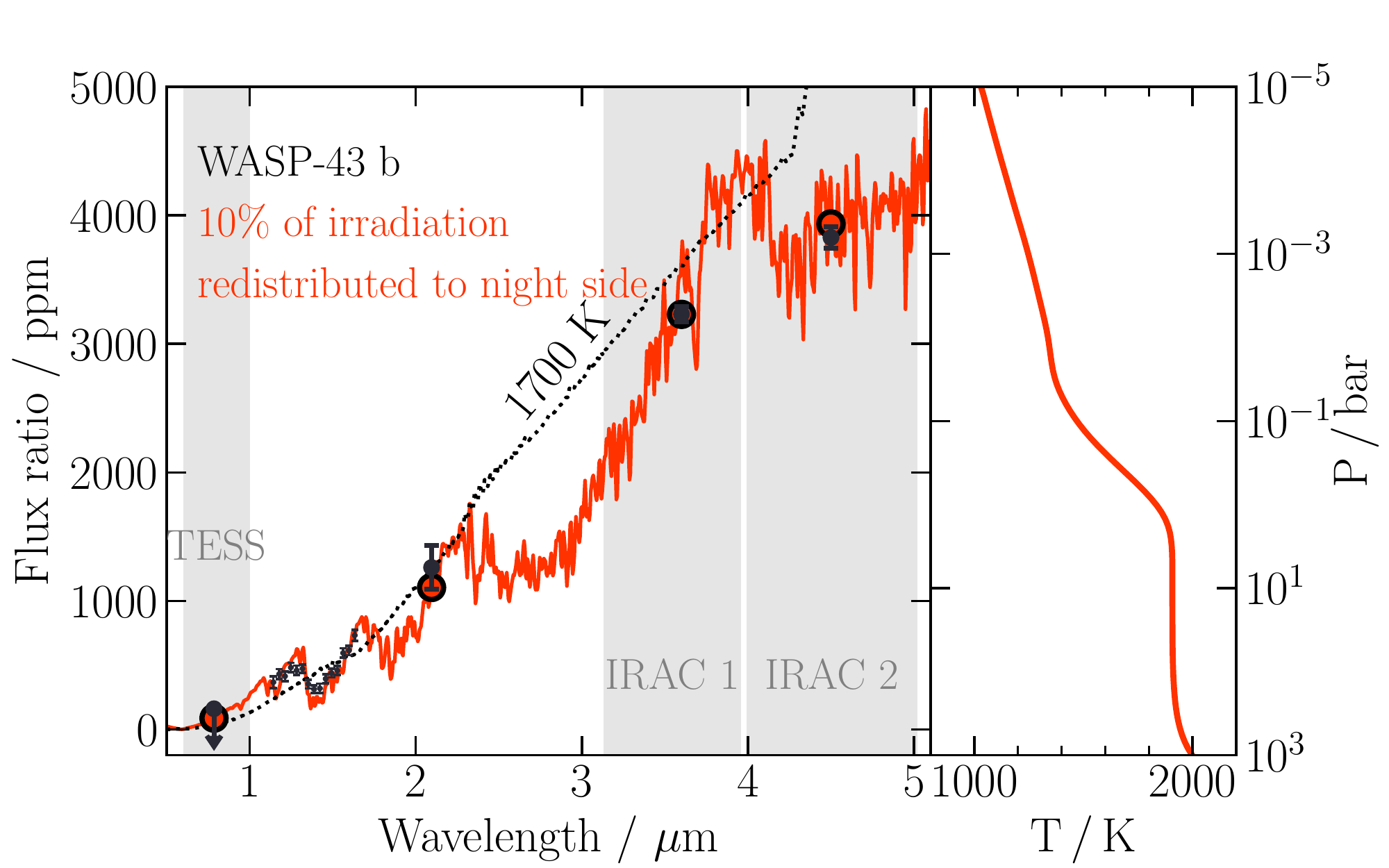}\\
    \includegraphics[width=0.5\textwidth]{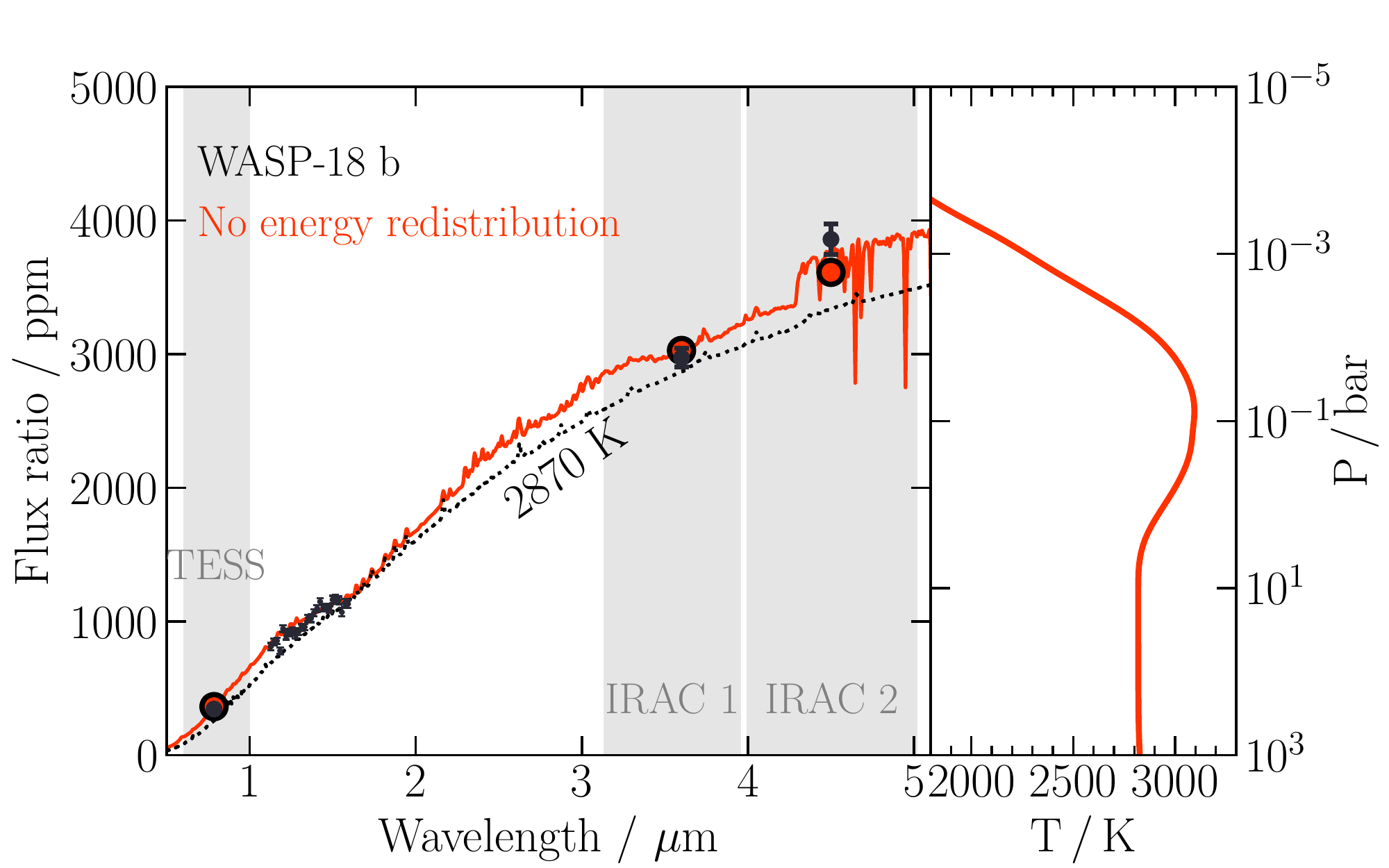}
        \caption{Self-consistent temperature pro{f}iles and thermal emission spectra for the dayside atmospheres of WASP-43\,b (top panel) and WASP-18\,b (bottom panel). \textit{TESS}, HAWK-I and \textit{Spitzer} observations (upper limits) are shown as black points and error bars (arrows), while the red circles show the binned model points. Note that the \emph{TESS} error bar for WASP-18\,b is smaller than the symbol size. The \textit{Spitzer} data for WASP-43\,b and WASP-18\,b are from \citet{Blecic2014} and \citet{Sheppard2017}, respectively. Small black points and error bars show \textit{HST}/WFC3 data for WASP-43\,b \citep{Kreidberg2014} and WASP-18\,b \citep{Sheppard2017}. The dashed black lines show blackbody spectra corresponding to the irradiation temperature, $T_{\rm irr}$, for each planet. $T_\mathrm{irr}=2^{-1/4}\sqrt{R_\star/a}\,T_{\rm e{f}f,\star}$ corresponds to the dayside temperature of the planet assuming no day-night energy redistribution and a Bond albedo of zero ($R_\star$, $a$, and $T_{\rm e{f}f,\star}$ de{f}ined as in Section \ref{sec:analysis}).}
    \label{fig:models}
\end{figure}

WASP-43\,b and WASP-18\,b represent opposite ends in temperature across the hot and ultra-hot Jupiter regimes. Therefore, they are ideal case studies for the comparative study of hot Jupiter atmospheres, including the presence of clouds and hazes. The \textit{TESS} and HAWK-I occultation depths we have derived for these planets (Section~\ref{sec:results}) provide constraints on the optical and near-infrared thermal emission/scattering of these planets. In this section, we therefore model the atmospheres of WASP-43\,b and WASP-18\,b in order to assess their potential atmospheric properties.

We self-consistently model the dayside atmospheres of WASP-43\,b and WASP-18\,b using the \textsc{genesis} atmospheric model \citep{Gandhi2017,Piette2020a}. \textsc{genesis} solves for the temperature pro{f}ile, thermal emission spectrum and chemical pro{f}ile of the atmosphere by calculating full, line-by-line radiative transfer under radiative-convective, thermodynamic, hydrostatic and thermochemical equilibrium. In particular, equilibrium chemical abundances are calculated using the \textsc{hsc chemistry} (version 8) software (see e.g., \citealt{Moriarty2014,Harrison2018,Piette2020a}). \textsc{hsc chemistry} minimizes the Gibbs' free energy of the system using the \textsc{gibbs} solver \citep{White1958}, given the atmospheric elemental abundances. These equilibrium chemistry calculations consider $>150$ chemical species (see \citealt{Piette2020a}). Of these, we consider atmospheric opacity due to the species known to dominate the H$_2$-rich atmospheres of hot Jupiters \citep{Burrows1999,Madhusudhan2016}: H$_2$O, CH$_4$, CO, CO$_2$, NH$_3$, HCN, C$_2$H$_2$, Na, K, TiO, VO and H$^-$, besides H$_2$ and He. We note that besides Na, K, TiO and VO, other atomic and molecular species such as Fe and AlO can also contribute  to the optical opacity and cause thermal inversions in hot Jupiter atmospheres \citep[e.g.,][]{Lothringer2018,Gandhi2019}. However, the optical data considered here (i.e., \textit{TESS} photometry) is only sensitive to the integrated optical {f}lux, and does not resolve spectral features due to individual species. We therefore use Na, K, TiO and VO as a proxy for the atmospheric optical opacity in these models, and {f}ind that we are able to explain the observations.

We calculate the absorption cross sections of these species as in \citet{Gandhi2017} using line lists from ExoMol, HITEMP and HITRAN (H$_2$O, CO and CO$_2$: \citealt{Rothman2010}, CH$_4$: \citealt{Yurchenko2013,Yurchenko2014a}, C$_2$H$_2$: \citealt{Rothman2013,Gordon2017}, NH$_3$: \citealt{Yurchenko2011}, HCN: \citealt{Harris2006,Barber2014}, TiO: \citealt{McKemmish2019}, VO: \citealt{McKemmish2016}, H$_2$-H$_2$ and H$_2$-He Collision-Induced Absorption: \citealt{Richard2012}). Na and K opacities are calculated as in \citet{Burrows2003} and \citet{Gandhi2017}, and H$^-$ bound-free and free-free cross sections are calculated using the prescriptions of \citet{Bell1987} and \citet{John1988} (see also \citealt{Arcangeli2018,parmentier2018b,Gandhi2020}). 

The free parameters in the atmospheric model are therefore the elemental abundances (explored here by changing the C/O ratio and metallicity), the incident irradiation, and the internal {f}lux. The incident irradiation on the dayside of a hot Jupiter can be varied by considering di{f}ferent e{f}{f}iciencies of energy redistribution, both on the day side and between the day and night sides \citep[see e.g.,][]{Burrows2008}, as described below. The internal {f}lux can be parameterised by a single temperature parameter ($T_{\rm int}$) and represents the {f}lux emanating from the planetary interior, e.g., as a remnant of the planet formation process. Given the relatively high irradiation levels of both WASP-43\,b and WASP-18\,b, the internal heat is not expected to noticeably a{f}fect the observable atmosphere. We therefore set $T_{\rm int}$ to a nominal value of 100\,K, similar to that of Jupiter. We explore physically plausible models for WASP-43\,b and WASP-18\,b in order to explain their observed \textit{TESS} and HAWK-I occultation depths (reported in this work) as well as existing \textit{Spitzer} IRAC dayside {f}luxes. The IRAC~1 and IRAC~2 data are obtained from \citet{Blecic2014} and \citet{Sheppard2017} for WASP-43\,b and WASP-18\,b, respectively. 

For WASP-43\,b, we {f}ind that an atmospheric model with solar metallicity and $\mathrm{C/O}=0.5$ is able to {f}it the observed \textit{TESS}, HAWK-I and \textit{Spitzer} data if 10\,per cent of the energy incident on the day side is transported to the night side, and energy redistribution is e{f}{f}icient on the dayside (top panel of Fig.~\ref{fig:models}). Using the notation of \citet{Burrows2008} and Equation~\ref{eq:teq}, this corresponds to a {f}lux distribution factor of $f=0.45$. Our model is in agreement with previous inferences of ine{f}{f}icient day-night energy redistribution from \textit{Spitzer} and TRAPPIST eclipse observations \citep{gillon2012,Blecic2014}. The strong day-night {f}lux contrast from \textit{Spitzer} phase curve constraints is also suggestive of ine{f}{f}icient day-night energy redistribution \citep{Stevenson2014,Stevenson2017}, though \citet{Stevenson2017} note that this contrast could also be caused by high-altitude nightside clouds. Our model {f}its the \textit{Spitzer} and HAWK-I NB2090 data within the $\sim1\sigma$ uncertainties, while models with more e{f}{f}icient day-night energy redistribution result in IRAC~1 and IRAC~2 brightness temperatures which are colder than what is observed. 

This atmospheric model for WASP-43\,b is dominated by H$_2$O and CO opacity, as expected for H$_2$-rich atmospheres at such temperatures \citep{Burrows1999,Madhusudhan2016}. The IRAC~1 and IRAC~2 bands probe H$_2$O and CO absorption features, respectively. Meanwhile, the \textit{TESS} and HAWK-I NB2090 bands probe the spectral continuum and therefore have a higher brightness temperature relative to the \textit{Spitzer} data. The model also agrees well with occultation data from the Hubble Space Telescope's Wide-Field Camera 3 (\textit{HST}/WFC3; \citealt{Kreidberg2014}), as shown in Fig.~\ref{fig:models}. We further note that the \textit{TESS} upper limit is consistent with pure thermal emission, without the need for re{f}lected light.

In the case of WASP-18\,b, we {f}ind that an atmospheric model with solar metallicity and $\mathrm{C/O}=1$ is able to {f}it the observed \textit{TESS} and \textit{Spitzer} data if there is no day-night energy redistribution and no energy redistribution on the dayside of the planet (i.e., instant re-radiation). This corresponds to a {f}lux distribution factor of $f=2/3$ \citep{Burrows2008} and is consistent with \textit{Spitzer} phase curve observations \citep{maxted}, while \citet{Arcangeli2019} infer a redistribution e{f}{f}iciency between uniform dayside redistribution ($f=0.5$) and instant re-radiation ($f=2/3$) from \textit{HST}/WFC3 phase curve observations. The model is shown in the bottom panel of Fig.~\ref{fig:models} and is able to {f}it the \textit{TESS} and \textit{Spitzer} observations within the $\sim2\sigma$ uncertainties.

This atmospheric model for WASP-18\,b is also broadly consistent with previous studies of its \textit{Spitzer} and \textit{HST}/WFC3 thermal emission observations \citep{Sheppard2017,Arcangeli2018,Gandhi2020}. For example, \citet{Sheppard2017} retrieve $\mathrm{C/O}=1$, while \citet{Gandhi2020} {f}ind evidence for sub-solar H$_2$O and super-solar CO (consistent with a high C/O ratio) and \citet{Arcangeli2018} derive a super-solar upper limit of $\mathrm{C/O}<0.85$. Furthermore, the atmospheric metallicity derived by \citet{Arcangeli2018} is consistent with solar values, though \citet{Sheppard2017} infer a super-solar metallicity and \citet{Gandhi2020} infer a metallicity between solar and super-solar, depending on the model assumptions and data used. We overplot the \textit{HST}/WFC3 data from \citet{Sheppard2017} in Fig.~\ref{fig:models} and {f}ind that these are in good agreement with our self-consistent model. We note that the photometric \textit{TESS} and \textit{Spitzer} data is not signi{f}icantly sensitive to the model C/O ratio, while the lack of H$_2$O absorption in the \textit{HST}/WFC3 data is better {f}it by a higher C/O ratio. Consistent with \citet{Sheppard2017}, \citet{Arcangeli2018}, and \citet{Gandhi2020}, we {f}ind that a thermal inversion is required to explain the \textit{Spitzer} data for WASP-18\,b. In particular, the IRAC~2 data point probes a CO emission feature and therefore has a higher brightness temperature than the \textit{TESS} and IRAC~1 observations. Furthermore, we {f}ind that with this model, the \textit{TESS} observation is readily explained by thermal emission alone, without the need for re{f}lected light.

\section{Conclusions}
In this work, we have presented constraints on the occultation depths and geometric albedos ($A_\mathrm{g}$) of {f}ive hot Jupiters using data from the \textit{TESS} space mission: WASP-18\,b, WASP-36\,b, WASP-43\,b, WASP-50\,b, and WASP-51\,b. We place the {f}irst constraints on the albedos of WASP-50\,b and WASP-51\,b, i.e., 3$\sigma$ upper limits of $A_\mathrm{g}<0.44$ and $A_\mathrm{g}<0.368$, respectively. For WASP-36\,b we place a 3$\sigma$ upper limit of $A_\mathrm{g}<0.286$, consistent with the previously published value of $0.16\pm0.16$ \citep{wong2020b}. We further con{f}irm the previous transit and occultation detections of WASP-18\,b with \textit{TESS}, and {f}ind a $3\sigma$ upper limit on the albedo, $A_\mathrm{g}<0.045$, consistent with the result of \citet{shporer}. We also place a 3$\sigma$ upper limit on the albedo of WASP-43\,b, $A_\mathrm{g}<0.154$, in the \textit{TESS} bandpass, consistent with the results of \citet{chen} and \citet{wong2020b}. 

Using data of the ground-based ESO VLT HAWK-I near-infrared instrument, we con{f}idently detect the occultation of WASP-43\,b. This data point is valuable for the modelling and characterisation of WASP-43\,b, and can be explained alongside existing \textit{Spitzer} data. Results of the same data set had been previously published in \citet{gillon2012}. We therefore used this data set as a benchmark to compare two di{f}ferent {f}itting methods and found out that the derived occultation depths agree within $\sim2\sigma$.

We use both the \textit{TESS} and HAWK-I data to place more detailed constraints on the atmospheres of two end-member hot Jupiters: WASP-43\,b and WASP-18\,b. To do this, we calculate self-consistent atmospheric models for each of these planets which explain the \textit{TESS}, HAWK-I and \textit{Spitzer} observations. As WASP-43\,b and WASP-18\,b represent opposite extremes in temperature, these data allow a comparative study of exoplanet atmospheres across the hot and ultra-hot Jupiter regimes. 

For both WASP-43\,b and WASP-18\,b, we {f}ind that ine{f}{f}icient energy redistribution is required to explain the data, though more so for WASP-18\,b. In particular, we {f}ind that 10-per cent day--night energy redistribution can explain the observations of WASP-43\,b, and no dayside or day-night energy redistribution (i.e., instant re-radiation) can explain the WASP-18\,b observations. This is consistent with the observed trend of lower energy redistribution e{f}{f}iciencies for highly-irradiated hot Jupiters \citep{Cowan2011}. Consistent with previous works (e.g., \citealt{Blecic2014,Stevenson2014,Sheppard2017,Arcangeli2018}), we {f}ind that a non-inverted (inverted) temperature pro{f}ile is required to explain the thermal emission spectrum of WASP-43\,b (WASP-18\,b). We further {f}ind that thermal emission alone is able to explain the observations, without the need for re{f}lected light resulting from clouds and/or hazes. Despite the extreme temperature contrast between  WASP-43\,b and WASP-18\,b, the data analysed in this work therefore do not suggest the presence of clouds and/or hazes on the dayside of either planet.

As the population of hot Jupiters with \textit{TESS} observations continues to grow, so too does our understanding of their atmospheric albedos. Furthermore, complementary infrared observations are essential in order to model and characterise these atmospheres in more detail. While optical occultation depths provide a measure of planetary geometric albedos, infrared spectra allow such albedos to be put into context, e.g., with atmospheric compositions and thermal pro{f}iles. Future more precise observations of albedos and thermal emission from hot Jupiters could enable population-level studies with joint constraints on the temperature structures, compositions, and sources of scattering in their atmospheres.

\section*{Acknowledgements}
M. Bla\v{z}ek, P. Kab\'{a}th, and M. Skarka would like to acknowledge a M\v{S}MT INTER-TRANSFER grant number LTT20015. A. Piette acknowledges {f}inancial support from the Science and Technology Facilities Council (STFC), UK, towards her doctoral programme. M. Skarka acknowledges the support from OP VVV Postdoc@MUNI (No. CZ.02.2.69/0.0/0.0/16$\_$027/0008360). C. C\'{a}ceres acknowledges support by ANID BASAL project FB210003 and ICM N\'ucleo Milenio de Formaci\'on Planetaria, NPF. Based on observations collected at the European Southern Observatory under ESO programme 086.C-0222(B). This paper includes data collected by the \textit{TESS} mission. Funding for the \textit{TESS} mission is provided by the NASA's Science Mission Directorate. We acknowledge \textsc{iraf}, distributed by the National Optical Astronomy Observatory, which is operated by the Association of Universities for Research in Astronomy (AURA) under a cooperative agreement with the National Science Foundation. We thank the editor and the anonymous reviewers for their helpful comments which improved quality of the article.

\section*{Data availability}
The data underlying this article are available from ESO archive via the query form (\url{http://archive.eso.org/eso/eso_archive_main.html}) under the programme speci{f}ied in Acknowledgements. This article also includes data collected by the \textit{TESS} mission, which are publicly available from the Mikulski Archive for Space Telescopes (MAST) (\url{https://archive.stsci.edu/}).


\bibliography{blazek_tess_albedos_arxiv}{}

\begin{thebibliography}{}
\expandafter\ifx\csname natexlab\endcsname\relax\def\natexlab#1{#1}\fi
\providecommand{\url}[1]{\href{#1}{#1}}
\providecommand{\dodoi}[1]{doi:~\href{http://doi.org/#1}{\nolinkurl{#1}}}
\providecommand{\doeprint}[1]{\href{http://ascl.net/#1}{\nolinkurl{http://ascl.net/#1}}}
\providecommand{\doarXiv}[1]{\href{https://arxiv.org/abs/#1}{\nolinkurl{https://arxiv.org/abs/#1}}}

\bibitem[{{Adams} {et~al.}(2021){Adams}, {Kataria}, {Batalha}, {Gao}, \&
  {Knutson}}]{Adams2021}
{Adams}, D., {Kataria}, T., {Batalha}, N., {Gao}, P., \& {Knutson}, H. 2021,
  arXiv e-prints, arXiv:2112.00041.
\newblock \doarXiv{2112.00041}

\bibitem[{{Anderson} {et~al.}(2010){Anderson}, {Gillon}, {Maxted}, {Barman},
  {Collier Cameron}, {Hellier}, {Queloz}, {Smalley}, \& {Triaud}}]{anderson}
{Anderson}, D.~R., {Gillon}, M., {Maxted}, P.~F.~L., {et~al.} 2010, \aap, 513,
  L3, \dodoi{10.1051/0004-6361/201014226}

\bibitem[{{Angerhausen} {et~al.}(2015){Angerhausen}, {DeLarme}, \&
  {Morse}}]{Angerhausen2015}
{Angerhausen}, D., {DeLarme}, E., \& {Morse}, J.~A. 2015, \pasp, 127, 1113,
  \dodoi{10.1086/683797}

\bibitem[{{Arcangeli} {et~al.}(2018){Arcangeli}, {D{\'e}sert}, {Line}, {Bean},
  {Parmentier}, {Stevenson}, {Kreidberg}, {Fortney}, {Mansfield}, \&
  {Showman}}]{Arcangeli2018}
{Arcangeli}, J., {D{\'e}sert}, J.-M., {Line}, M.~R., {et~al.} 2018, \apjl, 855,
  L30, \dodoi{10.3847/2041-8213/aab272}

\bibitem[{{Arcangeli} {et~al.}(2019){Arcangeli}, {D{\'e}sert}, {Parmentier},
  {Stevenson}, {Bean}, {Line}, {Kreidberg}, {Fortney}, \&
  {Showman}}]{Arcangeli2019}
{Arcangeli}, J., {D{\'e}sert}, J.-M., {Parmentier}, V., {et~al.} 2019, \aap,
  625, A136, \dodoi{10.1051/0004-6361/201834891}

\bibitem[{{Barber} {et~al.}(2014){Barber}, {Strange}, {Hill}, {Polyansky},
  {Mellau}, {Yurchenko}, \& {Tennyson}}]{Barber2014}
{Barber}, R.~J., {Strange}, J.~K., {Hill}, C., {et~al.} 2014, \mnras, 437,
  1828, \dodoi{10.1093/mnras/stt2011}

\bibitem[{{Batalha} {et~al.}(2011){Batalha}, {Borucki}, {Bryson}, {Buchhave},
  {Caldwell}, {Christensen-Dalsgaard}, {Ciardi}, {Dunham}, {Fressin},
  {Gautier}, {Gilliland}, {Haas}, {Howell}, {Jenkins}, {Kjeldsen}, {Koch},
  {Latham}, {Lissauer}, {Marcy}, {Rowe}, {Sasselov}, {Seager}, {Steffen},
  {Torres}, {Basri}, {Brown}, {Charbonneau}, {Christiansen}, {Clarke},
  {Cochran}, {Dupree}, {Fabrycky}, {Fischer}, {Ford}, {Fortney}, {Girouard},
  {Holman}, {Johnson}, {Isaacson}, {Klaus}, {Machalek}, {Moorehead},
  {Morehead}, {Ragozzine}, {Tenenbaum}, {Twicken}, {Quinn}, {VanCleve},
  {Walkowicz}, {Welsh}, {Devore}, \& {Gould}}]{batalha}
{Batalha}, N.~M., {Borucki}, W.~J., {Bryson}, S.~T., {et~al.} 2011, \apj, 729,
  27, \dodoi{10.1088/0004-637X/729/1/27}

\bibitem[{{Baxter} {et~al.}(2020){Baxter}, {D{\'e}sert}, {Parmentier}, {Line},
  {Fortney}, {Arcangeli}, {Bean}, {Todorov}, \& {Mansfield}}]{baxter2020}
{Baxter}, C., {D{\'e}sert}, J.-M., {Parmentier}, V., {et~al.} 2020, \aap, 639,
  A36, \dodoi{10.1051/0004-6361/201937394}

\bibitem[{{Beatty} {et~al.}(2020){Beatty}, {Wong}, {Fetherolf}, {Line},
  {Shporer}, {Stassun}, {Ricker}, {Seager}, {Winn}, {Jenkins}, {Louie},
  {Schlieder}, {Sha}, {Tenenbaum}, \& {Yahalomi}}]{beatty}
{Beatty}, T.~G., {Wong}, I., {Fetherolf}, T., {et~al.} 2020, \aj, 160, 211,
  \dodoi{10.3847/1538-3881/abb5aa}

\bibitem[{{Bell} \& {Berrington}(1987)}]{Bell1987}
{Bell}, K.~L., \& {Berrington}, K.~A. 1987, Journal of Physics B Atomic
  Molecular Physics, 20, 801, \dodoi{10.1088/0022-3700/20/4/019}

\bibitem[{{Bell} {et~al.}(2017){Bell}, {Nikolov}, {Cowan}, {Barstow}, {Barman},
  {Crossfield}, {Gibson}, {Evans}, {Sing}, {Knutson}, {Kataria}, {Lothringer},
  {Benneke}, \& {Schwartz}}]{bell}
{Bell}, T.~J., {Nikolov}, N., {Cowan}, N.~B., {et~al.} 2017, \apjl, 847, L2,
  \dodoi{10.3847/2041-8213/aa876c}

\bibitem[{{Blecic} {et~al.}(2014){Blecic}, {Harrington}, {Madhusudhan},
  {Stevenson}, {Hardy}, {Cubillos}, {Hardin}, {Bowman}, {Nymeyer}, {Anderson},
  {Hellier}, {Smith}, \& {Collier Cameron}}]{Blecic2014}
{Blecic}, J., {Harrington}, J., {Madhusudhan}, N., {et~al.} 2014, \apj, 781,
  116, \dodoi{10.1088/0004-637X/781/2/116}

\bibitem[{{Borucki} {et~al.}(2010){Borucki}, {Koch}, {Basri}, {Batalha},
  {Brown}, {Caldwell}, {Caldwell}, {Christensen-Dalsgaard}, {Cochran},
  {DeVore}, {Dunham}, {Dupree}, {Gautier}, {Geary}, {Gilliland}, {Gould},
  {Howell}, {Jenkins}, {Kondo}, {Latham}, {Marcy}, {Meibom}, {Kjeldsen},
  {Lissauer}, {Monet}, {Morrison}, {Sasselov}, {Tarter}, {Boss}, {Brownlee},
  {Owen}, {Buzasi}, {Charbonneau}, {Doyle}, {Fortney}, {Ford}, {Holman},
  {Seager}, {Steffen}, {Welsh}, {Rowe}, {Anderson}, {Buchhave}, {Ciardi},
  {Walkowicz}, {Sherry}, {Horch}, {Isaacson}, {Everett}, {Fischer}, {Torres},
  {Johnson}, {Endl}, {MacQueen}, {Bryson}, {Dotson}, {Haas}, {Kolodziejczak},
  {Van Cleve}, {Chandrasekaran}, {Twicken}, {Quintana}, {Clarke}, {Allen},
  {Li}, {Wu}, {Tenenbaum}, {Verner}, {Bruhweiler}, {Barnes}, \&
  {Prsa}}]{borucki}
{Borucki}, W.~J., {Koch}, D., {Basri}, G., {et~al.} 2010, Science, 327, 977,
  \dodoi{10.1126/science.1185402}

\bibitem[{{Bourrier} {et~al.}(2020){Bourrier}, {Kitzmann}, {Kuntzer},
  {Nascimbeni}, {Lendl}, {Lavie}, {Hoeijmakers}, {Pino}, {Ehrenreich}, {Heng},
  {Allart}, {Cegla}, {Dumusque}, {Melo}, {Astudillo-Defru}, {Caldwell},
  {Cretignier}, {Giles}, {Henze}, {Jenkins}, {Lovis}, {Murgas}, {Pepe},
  {Ricker}, {Rose}, {Seager}, {Segransan}, {Su{\'a}rez-Mascare{\~n}o}, {Udry},
  {Vanderspek}, \& {Wyttenbach}}]{bourrier2020}
{Bourrier}, V., {Kitzmann}, D., {Kuntzer}, T., {et~al.} 2020, \aap, 637, A36,
  \dodoi{10.1051/0004-6361/201936647}

\bibitem[{{Brandeker} {et~al.}(2022){Brandeker}, {Heng}, {Lendl}, {Patel},
  {Morris}, {Broeg}, {Guterman}, {Beck}, {Maxted}, {Demangeon}, {Delrez},
  {Demory}, {Kitzmann}, {Santos}, {Singh}, {Alibert}, {Alonso}, {Anglada},
  {B{\'a}rczy}, {Barrado y Navascues}, {Barros}, {Baumjohann}, {Beck}, {Benz},
  {Billot}, {Bonfils}, {Bruno}, {Cabrera}, {Charnoz}, {Collier Cameron},
  {Corral van Damme}, {Csizmadia}, {Davies}, {Deleuil}, {Deline}, {Ehrenreich},
  {Erikson}, {Farinato}, {Fortier}, {Fossati}, {Fridlund}, {Gandolfi},
  {Gillon}, {G{\"u}del}, {Hoyer}, {Isaak}, {Kiss}, {Laskar}, {Lecavelier des
  Etangs}, {Lovis}, {Luntzer}, {Magrin}, {Nascimbeni}, {Olofsson},
  {Ottensamer}, {Pagano}, {Pall{\'e}}, {Peter}, {Piotto}, {Pollacco}, {Queloz},
  {Ragazzoni}, {Rando}, {Rauer}, {Ribas}, {Scandariato}, {S{\'e}gransan},
  {Simon}, {Smith}, {Sousa}, {Steller}, {Szab{\'o}}, {Thomas}, {Udry}, {Van
  Grootel}, {Walton}, \& {Wolter}}]{brandeker2022}
{Brandeker}, A., {Heng}, K., {Lendl}, M., {et~al.} 2022, \aap, 659, L4,
  \dodoi{10.1051/0004-6361/202243082}

\bibitem[{{Burrows} {et~al.}(2008{\natexlab{a}}){Burrows}, {Budaj}, \&
  {Hubeny}}]{Burrows2008}
{Burrows}, A., {Budaj}, J., \& {Hubeny}, I. 2008{\natexlab{a}}, \apj, 678,
  1436, \dodoi{10.1086/533518}

\bibitem[{{Burrows} {et~al.}(2008{\natexlab{b}}){Burrows}, {Ibgui}, \&
  {Hubeny}}]{Burrows2008b}
{Burrows}, A., {Ibgui}, L., \& {Hubeny}, I. 2008{\natexlab{b}}, \apj, 682,
  1277, \dodoi{10.1086/589824}

\bibitem[{{Burrows} \& {Sharp}(1999)}]{Burrows1999}
{Burrows}, A., \& {Sharp}, C.~M. 1999, \apj, 512, 843, \dodoi{10.1086/306811}

\bibitem[{{Burrows} \& {Volobuyev}(2003)}]{Burrows2003}
{Burrows}, A., \& {Volobuyev}, M. 2003, \apj, 583, 985, \dodoi{10.1086/345412}

\bibitem[{{Casali} {et~al.}(2006){Casali}, {Pirard}, {Kissler-Patig},
  {Moorwood}, {Bedin}, {Biereichel}, {Delabre}, {Dorn}, {Finger}, {Gojak},
  {Huster}, {Jung}, {Koch}, {Lizon}, {Mehrgan}, {Pozna}, {Silber}, {Sokar}, \&
  {Stegmeier}}]{casali}
{Casali}, M., {Pirard}, J.-F., {Kissler-Patig}, M., {et~al.} 2006, in Society
  of Photo-Optical Instrumentation Engineers (SPIE) Conference Series, Vol.
  6269, Society of Photo-Optical Instrumentation Engineers (SPIE) Conference
  Series, ed. I.~S. {McLean} \& M.~{Iye}, 62690W, \dodoi{10.1117/12.670150}

\bibitem[{{Chakrabarty} \& {Sengupta}(2019)}]{chakrabarty}
{Chakrabarty}, A., \& {Sengupta}, S. 2019, \aj, 158, 39,
  \dodoi{10.3847/1538-3881/ab24dd}

\bibitem[{{Chen} {et~al.}(2014){Chen}, {van Boekel}, {Wang}, {Nikolov},
  {Fortney}, {Seemann}, {Wang}, {Mancini}, \& {Henning}}]{chen}
{Chen}, G., {van Boekel}, R., {Wang}, H., {et~al.} 2014, \aap, 563, A40,
  \dodoi{10.1051/0004-6361/201322740}

\bibitem[{{Cowan} \& {Agol}(2011)}]{Cowan2011}
{Cowan}, N.~B., \& {Agol}, E. 2011, \apj, 729, 54,
  \dodoi{10.1088/0004-637X/729/1/54}

\bibitem[{{Dai} {et~al.}(2017){Dai}, {Winn}, {Yu}, \& {Albrecht}}]{dai}
{Dai}, F., {Winn}, J.~N., {Yu}, L., \& {Albrecht}, S. 2017, \aj, 153, 40,
  \dodoi{10.3847/1538-3881/153/1/40}

\bibitem[{{Demory} {et~al.}(2011){Demory}, {Seager}, {Madhusudhan}, {Kjeldsen},
  {Christensen-Dalsgaard}, {Gillon}, {Rowe}, {Welsh}, {Adams}, {Dupree},
  {McCarthy}, {Kulesa}, {Borucki}, \& {Koch}}]{demory2011}
{Demory}, B.-O., {Seager}, S., {Madhusudhan}, N., {et~al.} 2011, \apjl, 735,
  L12, \dodoi{10.1088/2041-8205/735/1/L12}

\bibitem[{{Demory} {et~al.}(2013){Demory}, {de Wit}, {Lewis}, {Fortney},
  {Zsom}, {Seager}, {Knutson}, {Heng}, {Madhusudhan}, {Gillon}, {Barclay},
  {Desert}, {Parmentier}, \& {Cowan}}]{demory2013}
{Demory}, B.-O., {de Wit}, J., {Lewis}, N., {et~al.} 2013, \apjl, 776, L25,
  \dodoi{10.1088/2041-8205/776/2/L25}

\bibitem[{{Enoch} {et~al.}(2011){Enoch}, {Anderson}, {Barros}, {Brown},
  {Collier Cameron}, {Faedi}, {Gillon}, {H{\'e}brard}, {Lister}, {Queloz},
  {Santerne}, {Smalley}, {Street}, {Triaud}, {West}, {Bouchy}, {Bento},
  {Butters}, {Fossati}, {Haswell}, {Hellier}, {Holmes}, {Jehin}, {Lendl},
  {Maxted}, {McCormac}, {Miller}, {Moulds}, {Moutou}, {Norton}, {Parley},
  {Pepe}, {Pollacco}, {Segransan}, {Simpson}, {Skillen}, {Smith}, {Udry}, \&
  {Wheatley}}]{enoch}
{Enoch}, B., {Anderson}, D.~R., {Barros}, S.~C.~C., {et~al.} 2011, \aj, 142,
  86, \dodoi{10.1088/0004-6256/142/3/86}

\bibitem[{{Esteves} {et~al.}(2015){Esteves}, {De Mooij}, \&
  {Jayawardhana}}]{Esteves2015}
{Esteves}, L.~J., {De Mooij}, E. J.~W., \& {Jayawardhana}, R. 2015, \apj, 804,
  150, \dodoi{10.1088/0004-637X/804/2/150}

\bibitem[{{Evans} {et~al.}(2013){Evans}, {Pont}, {Sing}, {Aigrain}, {Barstow},
  {D{\'e}sert}, {Gibson}, {Heng}, {Knutson}, \& {Lecavelier des
  Etangs}}]{Evans2013}
{Evans}, T.~M., {Pont}, F., {Sing}, D.~K., {et~al.} 2013, \apjl, 772, L16,
  \dodoi{10.1088/2041-8205/772/2/L16}

\bibitem[{{Fraine} {et~al.}(2021){Fraine}, {Mayorga}, {Stevenson}, {Lewis},
  {Kataria}, {Bean}, {Bruno}, {Fortney}, {Kreidberg}, {Morley}, {Mouawad},
  {Todorov}, {Parmentier}, {Wakeford}, {Feng}, {Kilpatrick}, \&
  {Line}}]{fraine2021}
{Fraine}, J., {Mayorga}, L.~C., {Stevenson}, K.~B., {et~al.} 2021, \aj, 161,
  269, \dodoi{10.3847/1538-3881/abe8d6}

\bibitem[{{Gandhi} \& {Madhusudhan}(2017)}]{Gandhi2017}
{Gandhi}, S., \& {Madhusudhan}, N. 2017, \mnras, 472, 2334,
  \dodoi{10.1093/mnras/stx1601}

\bibitem[{{Gandhi} \& {Madhusudhan}(2019)}]{Gandhi2019}
---. 2019, \mnras, 485, 5817, \dodoi{10.1093/mnras/stz751}

\bibitem[{{Gandhi} {et~al.}(2020){Gandhi}, {Madhusudhan}, \&
  {Mandell}}]{Gandhi2020}
{Gandhi}, S., {Madhusudhan}, N., \& {Mandell}, A. 2020, \aj, 159, 232,
  \dodoi{10.3847/1538-3881/ab845e}

\bibitem[{{Gibson}(2014)}]{gibson2014}
{Gibson}, N.~P. 2014, \mnras, 445, 3401, \dodoi{10.1093/mnras/stu1975}

\bibitem[{{Gibson} {et~al.}(2013{\natexlab{a}}){Gibson}, {Aigrain}, {Barstow},
  {Evans}, {Fletcher}, \& {Irwin}}]{gibson2013a}
{Gibson}, N.~P., {Aigrain}, S., {Barstow}, J.~K., {et~al.} 2013{\natexlab{a}},
  \mnras, 428, 3680, \dodoi{10.1093/mnras/sts307}

\bibitem[{{Gibson} {et~al.}(2013{\natexlab{b}}){Gibson}, {Aigrain}, {Barstow},
  {Evans}, {Fletcher}, \& {Irwin}}]{gibson2013b}
---. 2013{\natexlab{b}}, \mnras, 436, 2974, \dodoi{10.1093/mnras/stt1783}

\bibitem[{{Gibson} {et~al.}(2012){Gibson}, {Aigrain}, {Roberts}, {Evans},
  {Osborne}, \& {Pont}}]{gibson2012}
{Gibson}, N.~P., {Aigrain}, S., {Roberts}, S., {et~al.} 2012, \mnras, 419,
  2683, \dodoi{10.1111/j.1365-2966.2011.19915.x}

\bibitem[{{Gibson} {et~al.}(2010){Gibson}, {Aigrain}, {Pollacco}, {Barros},
  {Hebb}, {Hrudkov{\'a}}, {Simpson}, {Skillen}, \& {West}}]{gibson2010}
{Gibson}, N.~P., {Aigrain}, S., {Pollacco}, D.~L., {et~al.} 2010, \mnras, 404,
  L114, \dodoi{10.1111/j.1745-3933.2010.00847.x}

\bibitem[{{Gillon} {et~al.}(2010){Gillon}, {Lanotte}, {Barman}, {Miller},
  {Demory}, {Deleuil}, {Montalb{\'a}n}, {Bouchy}, {Collier Cameron}, {Deeg},
  {Fortney}, {Fridlund}, {Harrington}, {Magain}, {Moutou}, {Queloz}, {Rauer},
  {Rouan}, \& {Schneider}}]{gillon2010}
{Gillon}, M., {Lanotte}, A.~A., {Barman}, T., {et~al.} 2010, \aap, 511, A3,
  \dodoi{10.1051/0004-6361/200913507}

\bibitem[{{Gillon} {et~al.}(2011){Gillon}, {Doyle}, {Lendl}, {Maxted},
  {Triaud}, {Anderson}, {Barros}, {Bento}, {Collier-Cameron}, {Enoch}, {Faedi},
  {Hellier}, {Jehin}, {Magain}, {Montalb{\'a}n}, {Pepe}, {Pollacco}, {Queloz},
  {Smalley}, {Segransan}, {Smith}, {Southworth}, {Udry}, {West}, \&
  {Wheatley}}]{gillon2011}
{Gillon}, M., {Doyle}, A.~P., {Lendl}, M., {et~al.} 2011, \aap, 533, A88,
  \dodoi{10.1051/0004-6361/201117198}

\bibitem[{{Gillon} {et~al.}(2012){Gillon}, {Triaud}, {Fortney}, {Demory},
  {Jehin}, {Lendl}, {Magain}, {Kabath}, {Queloz}, {Alonso}, {Anderson},
  {Collier Cameron}, {Fumel}, {Hebb}, {Hellier}, {Lanotte}, {Maxted},
  {Mowlavi}, \& {Smalley}}]{gillon2012}
{Gillon}, M., {Triaud}, A.~H.~M.~J., {Fortney}, J.~J., {et~al.} 2012, \aap,
  542, A4, \dodoi{10.1051/0004-6361/201218817}

\bibitem[{{Gordon} {et~al.}(2017){Gordon}, {Rothman}, {Hill}, {Kochanov},
  {Tan}, {Bernath}, {Birk}, {Boudon}, {Campargue}, {Chance}, {Drouin}, {Flaud},
  {Gamache}, {Hodges}, {Jacquemart}, {Perevalov}, {Perrin}, {Shine}, {Smith},
  {Tennyson}, {Toon}, {Tran}, {Tyuterev}, {Barbe}, {Cs{\'a}sz{\'a}r}, {Devi},
  {Furtenbacher}, {Harrison}, {Hartmann}, {Jolly}, {Johnson}, {Karman},
  {Kleiner}, {Kyuberis}, {Loos}, {Lyulin}, {Massie}, {Mikhailenko},
  {Moazzen-Ahmadi}, {M{\"u}ller}, {Naumenko}, {Nikitin}, {Polyansky}, {Rey},
  {Rotger}, {Sharpe}, {Sung}, {Starikova}, {Tashkun}, {Auwera}, {Wagner},
  {Wilzewski}, {Wcis{\l}o}, {Yu}, \& {Zak}}]{Gordon2017}
{Gordon}, I.~E., {Rothman}, L.~S., {Hill}, C., {et~al.} 2017, \jqsrt, 203, 3,
  \dodoi{10.1016/j.jqsrt.2017.06.038}

\bibitem[{{G{\"u}nther} \& {Daylan}(2019)}]{guenther2019}
{G{\"u}nther}, M.~N., \& {Daylan}, T. 2019, {Allesfitter: Flexible Star and
  Exoplanet Inference From Photometry and Radial Velocity}, Astrophysics Source
  Code Library.
\newblock \doeprint{1903.003}

\bibitem[{{G{\"u}nther} \& {Daylan}(2021)}]{guenther2021}
---. 2021, \apjs, 254, 13, \dodoi{10.3847/1538-4365/abe70e}

\bibitem[{{Harris} {et~al.}(2006){Harris}, {Tennyson}, {Kaminsky}, {Pavlenko},
  \& {Jones}}]{Harris2006}
{Harris}, G.~J., {Tennyson}, J., {Kaminsky}, B.~M., {Pavlenko}, Y.~V., \&
  {Jones}, H.~R.~A. 2006, \mnras, 367, 400,
  \dodoi{10.1111/j.1365-2966.2005.09960.x}

\bibitem[{{Harrison} {et~al.}(2018){Harrison}, {Bonsor}, \&
  {Madhusudhan}}]{Harrison2018}
{Harrison}, J. H.~D., {Bonsor}, A., \& {Madhusudhan}, N. 2018, \mnras, 479,
  3814, \dodoi{10.1093/mnras/sty1700}

\bibitem[{{Hellier} {et~al.}(2009){Hellier}, {Anderson}, {Collier Cameron},
  {Gillon}, {Hebb}, {Maxted}, {Queloz}, {Smalley}, {Triaud}, {West}, {Wilson},
  {Bentley}, {Enoch}, {Horne}, {Irwin}, {Lister}, {Mayor}, {Parley}, {Pepe},
  {Pollacco}, {Segransan}, {Udry}, \& {Wheatley}}]{hellier2009}
{Hellier}, C., {Anderson}, D.~R., {Collier Cameron}, A., {et~al.} 2009, \nat,
  460, 1098, \dodoi{10.1038/nature08245}

\bibitem[{{Hellier} {et~al.}(2011){Hellier}, {Anderson}, {Collier Cameron},
  {Gillon}, {Jehin}, {Lendl}, {Maxted}, {Pepe}, {Pollacco}, {Queloz},
  {S{\'e}gransan}, {Smalley}, {Smith}, {Southworth}, {Triaud}, {Udry}, \&
  {West}}]{hellier}
---. 2011, \aap, 535, L7, \dodoi{10.1051/0004-6361/201117081}

\bibitem[{{Heng} \& {Demory}(2013)}]{heng2013}
{Heng}, K., \& {Demory}, B.-O. 2013, \apj, 777, 100,
  \dodoi{10.1088/0004-637X/777/2/100}

\bibitem[{{Heng} {et~al.}(2021){Heng}, {Morris}, \& {Kitzmann}}]{Heng2021}
{Heng}, K., {Morris}, B.~M., \& {Kitzmann}, D. 2021, Nature Astronomy, 5, 1001,
  \dodoi{10.1038/s41550-021-01444-7}

\bibitem[{{Hoyer} {et~al.}(2016){Hoyer}, {Pall{\'e}}, {Dragomir}, \&
  {Murgas}}]{hoyer2016}
{Hoyer}, S., {Pall{\'e}}, E., {Dragomir}, D., \& {Murgas}, F. 2016, \aj, 151,
  137, \dodoi{10.3847/0004-6256/151/6/137}

\bibitem[{{Huber} {et~al.}(2017){Huber}, {Czesla}, \& {Schmitt}}]{huber}
{Huber}, K.~F., {Czesla}, S., \& {Schmitt}, J.~H.~M.~M. 2017, \aap, 597, A113,
  \dodoi{10.1051/0004-6361/201629699}

\bibitem[{Husser {et~al.}(2013)Husser, {Wende-von Berg}, Dreizler, Homeier,
  Reiners, Barman, \& Hauschildt}]{husser2013}
Husser, T.-O., {Wende-von Berg}, S., Dreizler, S., {et~al.} 2013, A{\&}A, 553,
  A6, \dodoi{10.1051/0004-6361/201219058}

\bibitem[{{Jenkins} {et~al.}(2016){Jenkins}, {Twicken}, {McCauliff},
  {Campbell}, {Sanderfer}, {Lung}, {Mansouri-Samani}, {Girouard}, {Tenenbaum},
  {Klaus}, {Smith}, {Caldwell}, {Chacon}, {Henze}, {Heiges}, {Latham},
  {Morgan}, {Swade}, {Rinehart}, \& {Vanderspek}}]{jenkins2016}
{Jenkins}, J.~M., {Twicken}, J.~D., {McCauliff}, S., {et~al.} 2016, in Society
  of Photo-Optical Instrumentation Engineers (SPIE) Conference Series, Vol.
  9913, Software and Cyberinfrastructure for Astronomy IV, ed. G.~{Chiozzi} \&
  J.~C. {Guzman}, 99133E, \dodoi{10.1117/12.2233418}

\bibitem[{{John}(1988)}]{John1988}
{John}, T.~L. 1988, \aap, 193, 189

\bibitem[{{Johnson} {et~al.}(2011){Johnson}, {Winn}, {Bakos}, {Hartman},
  {Morton}, {Torres}, {Kov{\'a}cs}, {Latham}, {Noyes}, {Sato}, {Esquerdo},
  {Fischer}, {Marcy}, {Howard}, {Buchhave}, {F{\H{u}}r{\'e}sz}, {Quinn},
  {B{\'e}ky}, {Sasselov}, {Stefanik}, {L{\'a}z{\'a}r}, {Papp}, \&
  {S{\'a}ri}}]{johnson}
{Johnson}, J.~A., {Winn}, J.~N., {Bakos}, G.~{\'A}., {et~al.} 2011, \apj, 735,
  24, \dodoi{10.1088/0004-637X/735/1/24}

\bibitem[{{Keating} \& {Cowan}(2017)}]{keating}
{Keating}, D., \& {Cowan}, N.~B. 2017, \apjl, 849, L5,
  \dodoi{10.3847/2041-8213/aa8b6b}

\bibitem[{{Kipping} \& {Spiegel}(2011)}]{kipping}
{Kipping}, D.~M., \& {Spiegel}, D.~S. 2011, \mnras, 417, L88,
  \dodoi{10.1111/j.1745-3933.2011.01127.x}

\bibitem[{{Kissler-Patig} {et~al.}(2008){Kissler-Patig}, {Pirard}, {Casali},
  {Moorwood}, {Ageorges}, {Alves de Oliveira}, {Baksai}, {Bedin}, {Bendek},
  {Biereichel}, {Delabre}, {Dorn}, {Esteves}, {Finger}, {Gojak}, {Huster},
  {Jung}, {Kiekebush}, {Klein}, {Koch}, {Lizon}, {Mehrgan}, {Petr-Gotzens},
  {Pritchard}, {Selman}, \& {Stegmeier}}]{kissler-patig}
{Kissler-Patig}, M., {Pirard}, J.~F., {Casali}, M., {et~al.} 2008, \aap, 491,
  941, \dodoi{10.1051/0004-6361:200809910}

\bibitem[{{Kreidberg} {et~al.}(2014){Kreidberg}, {Bean}, {D{\'e}sert}, {Line},
  {Fortney}, {Madhusudhan}, {Stevenson}, {Showman}, {Charbonneau},
  {McCullough}, {Seager}, {Burrows}, {Henry}, {Williamson}, {Kataria}, \&
  {Homeier}}]{Kreidberg2014}
{Kreidberg}, L., {Bean}, J.~L., {D{\'e}sert}, J.-M., {et~al.} 2014, \apjl, 793,
  L27, \dodoi{10.1088/2041-8205/793/2/L27}

\bibitem[{{Lothringer} {et~al.}(2018){Lothringer}, {Barman}, \&
  {Koskinen}}]{Lothringer2018}
{Lothringer}, J.~D., {Barman}, T., \& {Koskinen}, T. 2018, \apj, 866, 27,
  \dodoi{10.3847/1538-4357/aadd9e}

\bibitem[{{Maciejewski} {et~al.}(2016){Maciejewski}, {Dimitrov}, {Mancini},
  {Southworth}, {Ciceri}, {D'Ago}, {Bruni}, {Raetz}, {Nowak}, {Ohlert},
  {Puchalski}, {Saral}, {Derman}, {Petrucci}, {Jofre}, {Seeliger}, \&
  {Henning}}]{maciejewski}
{Maciejewski}, G., {Dimitrov}, D., {Mancini}, L., {et~al.} 2016, \actaa, 66,
  55.
\newblock \doarXiv{1603.03268}

\bibitem[{{Madhusudhan}(2019)}]{Madhusudhan2019}
{Madhusudhan}, N. 2019, \araa, 57, 617,
  \dodoi{10.1146/annurev-astro-081817-051846}

\bibitem[{{Madhusudhan} {et~al.}(2016){Madhusudhan}, {Ag{\'u}ndez}, {Moses}, \&
  {Hu}}]{Madhusudhan2016}
{Madhusudhan}, N., {Ag{\'u}ndez}, M., {Moses}, J.~I., \& {Hu}, Y. 2016, \ssr,
  205, 285, \dodoi{10.1007/s11214-016-0254-3}

\bibitem[{{Mallonn} {et~al.}(2019){Mallonn}, {K{\"o}hler}, {Alexoudi}, {von
  Essen}, {Granzer}, {Poppenhaeger}, \& {Strassmeier}}]{Mallonn2019}
{Mallonn}, M., {K{\"o}hler}, J., {Alexoudi}, X., {et~al.} 2019, \aap, 624, A62,
  \dodoi{10.1051/0004-6361/201935079}

\bibitem[{{Maxted} {et~al.}(2013){Maxted}, {Anderson}, {Doyle}, {Gillon},
  {Harrington}, {Iro}, {Jehin}, {Lafreni{\`e}re}, {Smalley}, \&
  {Southworth}}]{maxted}
{Maxted}, P.~F.~L., {Anderson}, D.~R., {Doyle}, A.~P., {et~al.} 2013, \mnras,
  428, 2645, \dodoi{10.1093/mnras/sts231}

\bibitem[{{McKemmish} {et~al.}(2019){McKemmish}, {Masseron}, {Hoeijmakers},
  {Perez-Mesa}, {Grimm}, {Yurchenko}, \& {Tennyson}}]{McKemmish2019}
{McKemmish}, L.~K., {Masseron}, T., {Hoeijmakers}, H.~J., {et~al.} 2019,
  \mnras, 488, 2836, \dodoi{10.1093/mnras/stz1818}

\bibitem[{{McKemmish} {et~al.}(2016){McKemmish}, {Yurchenko}, \&
  {Tennyson}}]{McKemmish2016}
{McKemmish}, L.~K., {Yurchenko}, S.~N., \& {Tennyson}, J. 2016, \mnras, 463,
  771, \dodoi{10.1093/mnras/stw1969}

\bibitem[{{Moriarty} {et~al.}(2014){Moriarty}, {Madhusudhan}, \&
  {Fischer}}]{Moriarty2014}
{Moriarty}, J., {Madhusudhan}, N., \& {Fischer}, D. 2014, \apj, 787, 81,
  \dodoi{10.1088/0004-637X/787/1/81}

\bibitem[{{Morley} {et~al.}(2013){Morley}, {Fortney}, {Kempton}, {Marley},
  {Visscher}, \& {Zahnle}}]{Morley2013}
{Morley}, C.~V., {Fortney}, J.~J., {Kempton}, E. M.~R., {et~al.} 2013, \apj,
  775, 33, \dodoi{10.1088/0004-637X/775/1/33}

\bibitem[{{Niraula} {et~al.}(2018){Niraula}, {Redfield}, {de Wit}, {Dai},
  {Mireles}, {Serindag}, \& {Shporer}}]{Niraula2018}
{Niraula}, P., {Redfield}, S., {de Wit}, J., {et~al.} 2018, arXiv e-prints,
  arXiv:1812.09227.
\newblock \doarXiv{1812.09227}

\bibitem[{{Parmentier} {et~al.}(2016){Parmentier}, {Fortney}, {Showman},
  {Morley}, \& {Marley}}]{Parmentier2016}
{Parmentier}, V., {Fortney}, J.~J., {Showman}, A.~P., {Morley}, C., \&
  {Marley}, M.~S. 2016, \apj, 828, 22, \dodoi{10.3847/0004-637X/828/1/22}

\bibitem[{{Parmentier} {et~al.}(2018){Parmentier}, {Line}, {Bean}, {Mansfield},
  {Kreidberg}, {Lupu}, {Visscher}, {D{\'e}sert}, {Fortney}, {Deleuil},
  {Arcangeli}, {Showman}, \& {Marley}}]{parmentier2018b}
{Parmentier}, V., {Line}, M.~R., {Bean}, J.~L., {et~al.} 2018, \aap, 617, A110,
  \dodoi{10.1051/0004-6361/201833059}

\bibitem[{Parviainen \& Aigrain(2015)}]{parviainen2015}
Parviainen, H., \& Aigrain, S. 2015, MNRAS, 453, 3821,
  \dodoi{10.1093/mnras/stv1857}

\bibitem[{{Piette} {et~al.}(2020){Piette}, {Madhusudhan}, {McKemmish},
  {Gandhi}, {Masseron}, \& {Welbanks}}]{Piette2020a}
{Piette}, A. A.~A., {Madhusudhan}, N., {McKemmish}, L.~K., {et~al.} 2020,
  \mnras, 496, 3870, \dodoi{10.1093/mnras/staa1592}

\bibitem[{{Pirard} {et~al.}(2004){Pirard}, {Kissler-Patig}, {Moorwood},
  {Biereichel}, {Delabre}, {Dorn}, {Finger}, {Gojak}, {Huster}, {Jung}, {Koch},
  {Le Louarn}, {Lizon}, {Mehrgan}, {Pozna}, {Silber}, {Sokar}, \&
  {Stegmeier}}]{pirard}
{Pirard}, J.-F., {Kissler-Patig}, M., {Moorwood}, A., {et~al.} 2004, in Society
  of Photo-Optical Instrumentation Engineers (SPIE) Conference Series, Vol.
  5492, Ground-based Instrumentation for Astronomy, ed. A.~F.~M. {Moorwood} \&
  M.~{Iye}, 1763--1772, \dodoi{10.1117/12.578293}

\bibitem[{{Pollacco} {et~al.}(2006){Pollacco}, {Skillen}, {Collier Cameron},
  {Christian}, {Hellier}, {Irwin}, {Lister}, {Street}, {West}, {Anderson},
  {Clarkson}, {Deeg}, {Enoch}, {Evans}, {Fitzsimmons}, {Haswell}, {Hodgkin},
  {Horne}, {Kane}, {Keenan}, {Maxted}, {Norton}, {Osborne}, {Parley}, {Ryans},
  {Smalley}, {Wheatley}, \& {Wilson}}]{pollacco}
{Pollacco}, D.~L., {Skillen}, I., {Collier Cameron}, A., {et~al.} 2006, \pasp,
  118, 1407, \dodoi{10.1086/508556}

\bibitem[{{Richard} {et~al.}(2012){Richard}, {Gordon}, {Rothman}, {Abel},
  {Frommhold}, {Gustafsson}, {Hartmann}, {Hermans}, {Lafferty}, {Orton},
  {Smith}, \& {Tran}}]{Richard2012}
{Richard}, C., {Gordon}, I.~E., {Rothman}, L.~S., {et~al.} 2012, \jqsrt, 113,
  1276, \dodoi{10.1016/j.jqsrt.2011.11.004}

\bibitem[{{Ricker} {et~al.}(2015){Ricker}, {Winn}, {Vanderspek}, {Latham},
  {Bakos}, {Bean}, {Berta-Thompson}, {Brown}, {Buchhave}, {Butler}, {Butler},
  {Chaplin}, {Charbonneau}, {Christensen-Dalsgaard}, {Clampin}, {Deming},
  {Doty}, {De Lee}, {Dressing}, {Dunham}, {Endl}, {Fressin}, {Ge}, {Henning},
  {Holman}, {Howard}, {Ida}, {Jenkins}, {Jernigan}, {Johnson}, {Kaltenegger},
  {Kawai}, {Kjeldsen}, {Laughlin}, {Levine}, {Lin}, {Lissauer}, {MacQueen},
  {Marcy}, {McCullough}, {Morton}, {Narita}, {Paegert}, {Palle}, {Pepe},
  {Pepper}, {Quirrenbach}, {Rinehart}, {Sasselov}, {Sato}, {Seager},
  {Sozzetti}, {Stassun}, {Sullivan}, {Szentgyorgyi}, {Torres}, {Udry}, \&
  {Villasenor}}]{ricker2015}
{Ricker}, G.~R., {Winn}, J.~N., {Vanderspek}, R., {et~al.} 2015, Journal of
  Astronomical Telescopes, Instruments, and Systems, 1, 014003,
  \dodoi{10.1117/1.JATIS.1.1.014003}

\bibitem[{{Rothman} {et~al.}(2010){Rothman}, {Gordon}, {Barber}, {Dothe},
  {Gamache}, {Goldman}, {Perevalov}, {Tashkun}, \& {Tennyson}}]{Rothman2010}
{Rothman}, L.~S., {Gordon}, I.~E., {Barber}, R.~J., {et~al.} 2010, \jqsrt, 111,
  2139, \dodoi{10.1016/j.jqsrt.2010.05.001}

\bibitem[{{Rothman} {et~al.}(2013){Rothman}, {Gordon}, {Babikov}, {Barbe},
  {Chris Benner}, {Bernath}, {Birk}, {Bizzocchi}, {Boudon}, {Brown},
  {Campargue}, {Chance}, {Cohen}, {Coudert}, {Devi}, {Drouin}, {Fayt}, {Flaud},
  {Gamache}, {Harrison}, {Hartmann}, {Hill}, {Hodges}, {Jacquemart}, {Jolly},
  {Lamouroux}, {Le Roy}, {Li}, {Long}, {Lyulin}, {Mackie}, {Massie},
  {Mikhailenko}, {M{\"u}ller}, {Naumenko}, {Nikitin}, {Orphal}, {Perevalov},
  {Perrin}, {Polovtseva}, {Richard}, {Smith}, {Starikova}, {Sung}, {Tashkun},
  {Tennyson}, {Toon}, {Tyuterev}, \& {Wagner}}]{Rothman2013}
{Rothman}, L.~S., {Gordon}, I.~E., {Babikov}, Y., {et~al.} 2013, \jqsrt, 130,
  4, \dodoi{10.1016/j.jqsrt.2013.07.002}

\bibitem[{{Saeed} {et~al.}(2022){Saeed}, {Goderya}, \& {Chishtie}}]{saeed}
{Saeed}, M.~I., {Goderya}, S.~N., \& {Chishtie}, F.~A. 2022, \na, 91, 101680,
  \dodoi{10.1016/j.newast.2021.101680}

\bibitem[{{Saha} {et~al.}(2021){Saha}, {Chakrabarty}, \& {Sengupta}}]{saha}
{Saha}, S., {Chakrabarty}, A., \& {Sengupta}, S. 2021, \aj, 162, 18,
  \dodoi{10.3847/1538-3881/ac01dd}

\bibitem[{{Sheppard} {et~al.}(2017){Sheppard}, {Mandell}, {Tamburo}, {Gandhi},
  {Pinhas}, {Madhusudhan}, \& {Deming}}]{Sheppard2017}
{Sheppard}, K.~B., {Mandell}, A.~M., {Tamburo}, P., {et~al.} 2017, \apjl, 850,
  L32, \dodoi{10.3847/2041-8213/aa9ae9}

\bibitem[{{Shporer} \& {Hu}(2015)}]{Shporer2015}
{Shporer}, A., \& {Hu}, R. 2015, \aj, 150, 112,
  \dodoi{10.1088/0004-6256/150/4/112}

\bibitem[{{Shporer} {et~al.}(2019){Shporer}, {Wong}, {Huang}, {Line},
  {Stassun}, {Fetherolf}, {Kane}, {Bouma}, {Daylan}, {G{\"u}enther}, {Ricker},
  {Latham}, {Vanderspek}, {Seager}, {Winn}, {Jenkins}, {Glidden},
  {Berta-Thompson}, {Ting}, {Li}, \& {Haworth}}]{shporer}
{Shporer}, A., {Wong}, I., {Huang}, C.~X., {et~al.} 2019, \aj, 157, 178,
  \dodoi{10.3847/1538-3881/ab0f96}

\bibitem[{{Siebenmorgen} {et~al.}(2011){Siebenmorgen}, {Carraro}, {Valenti},
  {Petr-Gotzens}, {Brammer}, {Garcia}, \& {Casali}}]{siebenmorgen}
{Siebenmorgen}, R., {Carraro}, G., {Valenti}, E., {et~al.} 2011, The Messenger,
  144, 9

\bibitem[{{Smith} {et~al.}(2012{\natexlab{a}}){Smith}, {Anderson}, {Collier
  Cameron}, {Gillon}, {Hellier}, {Lendl}, {Maxted}, {Queloz}, {Smalley},
  {Triaud}, {West}, {Barros}, {Jehin}, {Pepe}, {Pollacco}, {Segransan},
  {Southworth}, {Street}, \& {Udry}}]{smith}
{Smith}, A.~M.~S., {Anderson}, D.~R., {Collier Cameron}, A., {et~al.}
  2012{\natexlab{a}}, \aj, 143, 81, \dodoi{10.1088/0004-6256/143/4/81}

\bibitem[{{Smith} {et~al.}(2012{\natexlab{b}}){Smith}, {Stumpe}, {Van Cleve},
  {Jenkins}, {Barclay}, {Fanelli}, {Girouard}, {Kolodziejczak}, {McCauliff},
  {Morris}, \& {Twicken}}]{smith2012}
{Smith}, J.~C., {Stumpe}, M.~C., {Van Cleve}, J.~E., {et~al.}
  2012{\natexlab{b}}, \pasp, 124, 1000, \dodoi{10.1086/667697}

\bibitem[{{Southworth} {et~al.}(2009){Southworth}, {Hinse}, {Dominik},
  {Glitrup}, {J{\o}rgensen}, {Liebig}, {Mathiasen}, {Anderson}, {Bozza},
  {Browne}, {Burgdorf}, {Calchi Novati}, {Dreizler}, {Finet}, {Harps{\o}e},
  {Hessman}, {Hundertmark}, {Maier}, {Mancini}, {Maxted}, {Rahvar}, {Ricci},
  {Scarpetta}, {Skottfelt}, {Snodgrass}, {Surdej}, \& {Zimmer}}]{southworth}
{Southworth}, J., {Hinse}, T.~C., {Dominik}, M., {et~al.} 2009, \apj, 707, 167,
  \dodoi{10.1088/0004-637X/707/1/167}

\bibitem[{{Stevenson} {et~al.}(2014){Stevenson}, {D{\'e}sert}, {Line}, {Bean},
  {Fortney}, {Showman}, {Kataria}, {Kreidberg}, {McCullough}, {Henry},
  {Charbonneau}, {Burrows}, {Seager}, {Madhusudhan}, {Williamson}, \&
  {Homeier}}]{Stevenson2014}
{Stevenson}, K.~B., {D{\'e}sert}, J.-M., {Line}, M.~R., {et~al.} 2014, Science,
  346, 838, \dodoi{10.1126/science.1256758}

\bibitem[{{Stevenson} {et~al.}(2017){Stevenson}, {Line}, {Bean}, {D{\'e}sert},
  {Fortney}, {Showman}, {Kataria}, {Kreidberg}, \& {Feng}}]{Stevenson2017}
{Stevenson}, K.~B., {Line}, M.~R., {Bean}, J.~L., {et~al.} 2017, \aj, 153, 68,
  \dodoi{10.3847/1538-3881/153/2/68}

\bibitem[{{Stumpe} {et~al.}(2012){Stumpe}, {Smith}, {Van Cleve}, {Twicken},
  {Barclay}, {Fanelli}, {Girouard}, {Jenkins}, {Kolodziejczak}, {McCauliff}, \&
  {Morris}}]{stumpe2012}
{Stumpe}, M.~C., {Smith}, J.~C., {Van Cleve}, J.~E., {et~al.} 2012, \pasp, 124,
  985, \dodoi{10.1086/667698}

\bibitem[{{Tody}(1986)}]{tody1}
{Tody}, D. 1986, in Society of Photo-Optical Instrumentation Engineers (SPIE)
  Conference Series, Vol. 627, \procspie, ed. D.~L. {Crawford}, 733,
  \dodoi{10.1117/12.968154}

\bibitem[{{Tody}(1993)}]{tody2}
{Tody}, D. 1993, in Astronomical Society of the Pacific Conference Series,
  Vol.~52, Astronomical Data Analysis Software and Systems II, ed. R.~J.
  {Hanisch}, R.~J.~V. {Brissenden}, \& J.~{Barnes}, 173

\bibitem[{{Tregloan-Reed} \& {Southworth}(2013)}]{tregloan-reed}
{Tregloan-Reed}, J., \& {Southworth}, J. 2013, \mnras, 431, 966,
  \dodoi{10.1093/mnras/stt227}

\bibitem[{White {et~al.}(1958)White, Johnson, \& Dantzig}]{White1958}
White, W.~B., Johnson, S.~M., \& Dantzig, G.~B. 1958, The Journal of Chemical
  Physics, 28, 751, \dodoi{10.1063/1.1744264}

\bibitem[{{Winn}(2010)}]{winn_book}
{Winn}, J.~N. 2010, {Exoplanet Transits and Occultations} ({University of
  Arizona space science series}), 55--77

\bibitem[{{Wong} {et~al.}(2020{\natexlab{a}}){Wong}, {Shporer}, {Kitzmann},
  {Morris}, {Heng}, {Hoeijmakers}, {Demory}, {Ahlers}, {Mansfield}, {Bean},
  {Daylan}, {Fetherolf}, {Rodriguez}, {Benneke}, {Ricker}, {Latham},
  {Vanderspek}, {Seager}, {Winn}, {Jenkins}, {Burke}, {Christiansen}, {Essack},
  {Rose}, {Smith}, {Tenenbaum}, \& {Yahalomi}}]{wong2020b}
{Wong}, I., {Shporer}, A., {Kitzmann}, D., {et~al.} 2020{\natexlab{a}}, \aj,
  160, 88, \dodoi{10.3847/1538-3881/aba2cb}

\bibitem[{{Wong} {et~al.}(2020{\natexlab{b}}){Wong}, {Shporer}, {Daylan},
  {Benneke}, {Fetherolf}, {Kane}, {Ricker}, {Vanderspek}, {Latham}, {Winn},
  {Jenkins}, {Boyd}, {Glidden}, {Goeke}, {Sha}, {Ting}, \& {Yahalomi}}]{wong}
{Wong}, I., {Shporer}, A., {Daylan}, T., {et~al.} 2020{\natexlab{b}}, \aj, 160,
  155, \dodoi{10.3847/1538-3881/ababad}

\bibitem[{{Yurchenko} {et~al.}(2011){Yurchenko}, {Barber}, \&
  {Tennyson}}]{Yurchenko2011}
{Yurchenko}, S.~N., {Barber}, R.~J., \& {Tennyson}, J. 2011, \mnras, 413, 1828,
  \dodoi{10.1111/j.1365-2966.2011.18261.x}

\bibitem[{{Yurchenko} \& {Tennyson}(2014)}]{Yurchenko2014a}
{Yurchenko}, S.~N., \& {Tennyson}, J. 2014, \mnras, 440, 1649,
  \dodoi{10.1093/mnras/stu326}

\bibitem[{{Yurchenko} {et~al.}(2013){Yurchenko}, {Tennyson}, {Barber}, \&
  {Thiel}}]{Yurchenko2013}
{Yurchenko}, S.~N., {Tennyson}, J., {Barber}, R.~J., \& {Thiel}, W. 2013,
  Journal of Molecular Spectroscopy, 291, 69, \dodoi{10.1016/j.jms.2013.05.014}

\bibitem[{{Zhou} {et~al.}(2015){Zhou}, {Bayliss}, {Kedziora-Chudczer},
  {Tinney}, {Bailey}, {Salter}, \& {Rodriguez}}]{zhou}
{Zhou}, G., {Bayliss}, D.~D.~R., {Kedziora-Chudczer}, L., {et~al.} 2015,
  \mnras, 454, 3002, \dodoi{10.1093/mnras/stv2138}

\end{thebibliography}
\bibliographystyle{aasjournal}



\end{document}